\documentclass[
    aps
    prl,
    preprint,
    groupedaddress,
    amsmath,amssymb
    ]{revtex4-1}

\usepackage[utf8]{inputenc}

\usepackage[USenglish]{babel}			 

\usepackage{multirow}
\usepackage{ragged2e}

\usepackage{amsmath}
\usepackage{amssymb}
\usepackage{appendix}
\usepackage{bm}
\usepackage{booktabs}
\usepackage{graphicx}
\usepackage{hyperref}
\usepackage{multirow}
\usepackage{mwe}
\usepackage{siunitx}
\usepackage{subcaption}
\usepackage{tablefootnote}
\usepackage{tabularx}
\usepackage{tcolorbox}
\usepackage{tikz}

\usetikzlibrary{arrows,decorations,backgrounds,patterns,matrix,shapes,fit,calc,shadows,plotmarks,pgfplots.groupplots}

\newcommand{\J}[2]{J_{#1}\left(#2\right)}
\newcommand{\Y}[2]{Y_{#1}\left(#2\right)}

\newcommand{\req}[1]{Eq.~(\ref{#1})}
\newcommand{\rfig}[1]{Fig.~\ref{#1}}
\newcommand{\rtable}[1]{Table~\ref{#1}}

\newcommand{\ExB}{$\vec{\,\mathrm{E}}\times\vec{\,\mathrm{B}}$ }

\DeclareMathOperator*{\argmin}{arg\,min}

\begin{document}

\makeatletter
\renewcommand\@dotsep{10000}
\makeatother

\title{Space-charge distortion of transverse profiles measured by electron-based Ionization Profile Monitors and correction methods}
\author{D. Vilsmeier}\email{d.vilsmeier@gsi.de}
\affiliation{Johann Wolfgang Goethe-University Frankfurt, 60323 Frankfurt am Main, Germany}
\author{M. Sapinski}
\author{R. Singh}
\affiliation{GSI Helmholtz Centre for Heavy Ion Research, 64291 Darmstadt, Planckstr. 1, Germany}
\date{\today}

\begin{abstract}
Measurements of transverse profiles using Ionization Profile Monitors (IPMs) for high brightness beams are affected by the electromagnetic field of the beam. This interaction may cause a distortion of the measured profile shape despite strong external magnetic field applied to impose limits on the transverse movement of electrons. The mechanisms leading to this distortion are discussed in detail. The distortion itself is described by means of analytic calculations for simplified beam distributions and a full simulation model for realistic distributions. Simple relation for minimum magnetic field scaling with beam parameters for avoiding profile distortions is presented. Further, application of machine learning algorithms to the problem of reconstructing the actual beam profile from distorted measured profile is presented.
The obtained results show good agreement for tests on simulation data.
The performance of these algorithms indicate that they could be very useful for operations of IPMs on high brightness beams or IPMs with weak magnetic field.
\end{abstract}

\maketitle

\section{Introduction}

\begin{figure}[htbp]
    \centering
    \includegraphics[width=8.6cm]{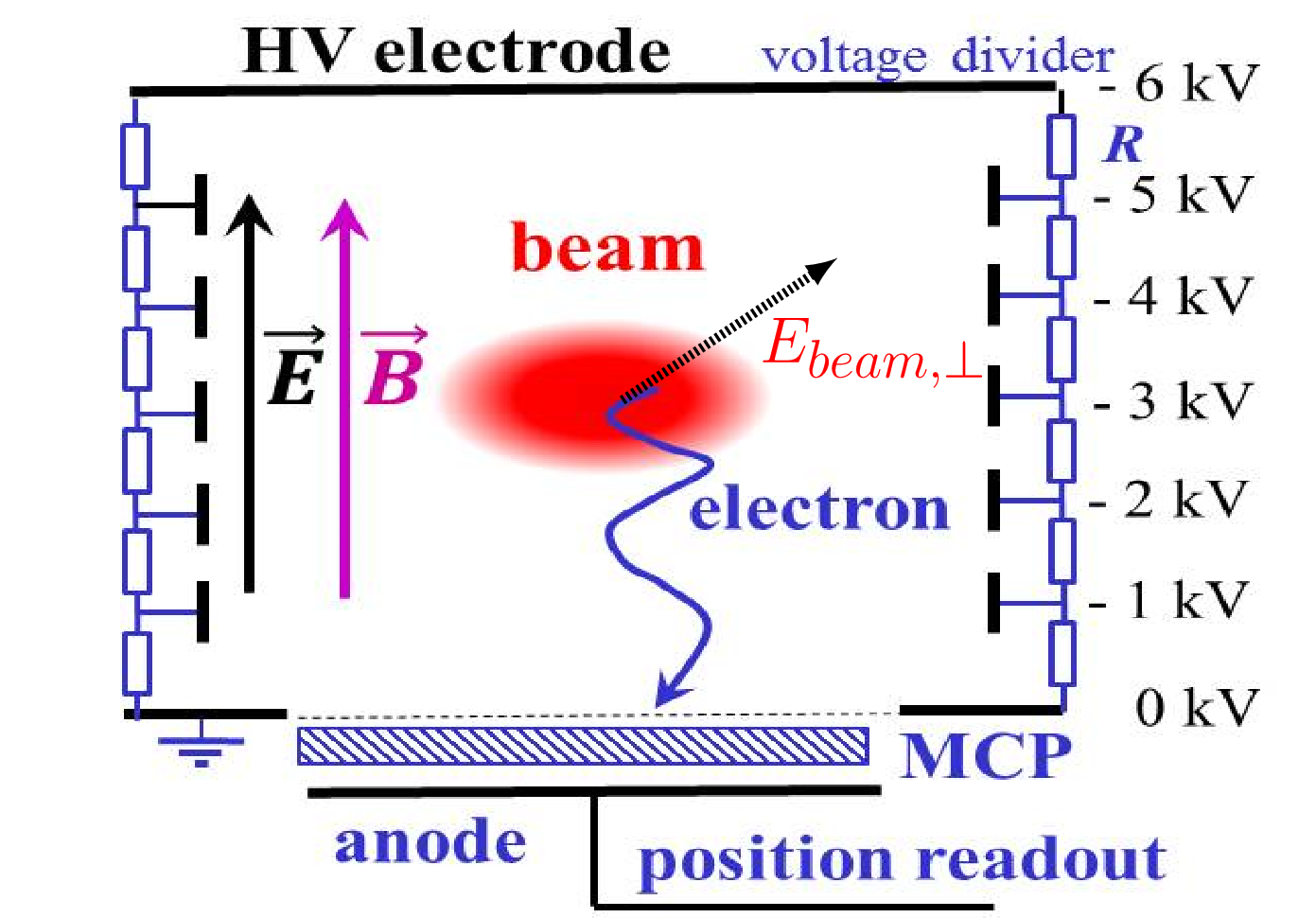}
    \caption[IPM sketch]{Sketch of an IPM together with read-out system. The electric and magnetic guiding fields are aligned, perpendicular to the beam direction as well as to the read-out system~\cite{ForckJUAS}.}
    \label{fig:01:ipm-sketch}
\end{figure}

The principle of beam profile measurement using Ionization Profile Monitors (IPMs) is seemingly very simple. The beam particles ionize the residual gas. The products of the ionization - electrons or ions - are extracted towards a position-sensitive detector using the guiding electric field (also referred to as 'clearing' or 'external' field) provided by electrodes. The distribution of the particles on the detector ideally corresponds to a projection of the transverse distribution of the beam density. 
This simple and straightforward principle, illustrated in \rfig{fig:01:ipm-sketch}, is non-destructive for the beam and thus an appealing application for synchrotrons and storage rings.
In this context IPMs are installed in many hadron accelerators~\cite{GSIworkshop2017} and even investigated for usage 
on electron machines~\cite{Miessner2011}.

However already for medium-intensity beams the movement of electrons or ions is easily affected by the transient beam electric field.
Because of their lower mass, electrons are removed much faster from the influence of bunch fields, however the effect of these fields on their 
trajectories is proportionally larger.
Usually ion-based devices are preferred due to simpler assembly (no need of ion-trapping~\cite{SatouIPAC2018}), smaller impact of fringe fields of neighbouring magnets and lack of background signal from electrons generated from various other sources~\cite{SatouIBIC2017}.
However the influence of beam fields plays a major role for high-brightness beams and thus electron-based IPMs are preferred over ion-based devices, mainly because the electron movement can be easily confined to small gyroradius by applying external magnetic field.  
Another argument in favor of electron-based devices is the fact that they are more suitable for time-resolved measurements because the electrons
have much smaller time-of-flights towards the detector.

Distortions of the measured beam profile using ion-based IPMs are either mitigated by increasing the electric guiding field or corrected by empirical mathematical models~\cite{DeLuca,Thern,Graves,Amundson}. For operation of electron-based IPMs with magnetic field it was first proposed to tune the magnetic field strength such that the electrons perform exactly one gyration between the beam and the detector~\cite{Hornsta1970}. However this method does not work for high-brightness beams, in which electrons get a significant momentum transfer not only transverse to the detector plane but also in the direction towards the detector. Therefore increasing the magnetic field in order to reduce the magnitude of gyroradius is the typical counteraction against the distortion of the measured beam profiles. This is also because profile correction methods, similar to the ones developed for ion-based IPMs, do not exist for electron-based devices. One reason for that is the more complex particle movement in the presence of a combination of electric and magnetic fields. This movement and therefore the profile distortion, strongly depend on the beam fields and hence on bunch charge, transverse and longitudinal bunch shape as well as bunch spacing.

The first beam space-charge induced effects in an electron-based IPM operated with magnetic guiding field were observed in the Large Hadron Collider (LHC) at beam energies of \SI{4}{\tera\electronvolt} and magnetic guiding field of \SI{0.2}{\tesla}~\cite{Sapinski-first-exp-IPM}. Following studies have analyzed this phenomenon and estimated a magnetic field strength of \SI{1}{\tesla} required to suppress profile distortions~\cite{Patecki-Thesis, Sapinski-HB-2014, Vilsmeier-Thesis}. However such large field strengths pose a technological as well as financial challenge, especially because of the relatively large gap between the poles needed to fit the detector vacuum chamber.

Besides beam space-charge induced profile distortion there are various other phenomena which can have an effect on measured profiles. These are briefly discussed below however not addressed any further throughout this study:
\begin{enumerate}
\item Ionization by bunch field. At extreme field strengths the gas ionization cross-section is modified due to the Stark shift of energy levels and at even higher fields the gas gets ionized by the collective bunch electric field~\cite{Keldysh}, rather than by interaction with single beam particles.
\item Burnout of the rest gas. Under some conditions, the ionization rate of the rest gas can be higher than the replenishment of gas in the volume of the beam leading to non-proportionality between beam density and amount of ionization events. This can be counterbalanced by, for instance, using gas jet~\cite{ZhangIBIC2015}.
\item Guiding field non-uniformity. If the electric and magnetic guiding fields are non-uniform, it will additionally impact the resulting particle trajectories~\cite{WilcoxIBIC2016}. Also if the guiding fields are not perfectly aligned with respect to each other this introduces an additional disturbance.
\item Gas ionization from synchrotron radiation and the influence of wake fields on particle trajectories may constitute additional sources of profile distortion. However these effects have not been observed or studied up to now.
\end{enumerate}

\begin{figure}[htbp]
    \centering
    \includegraphics[width=8.6cm]{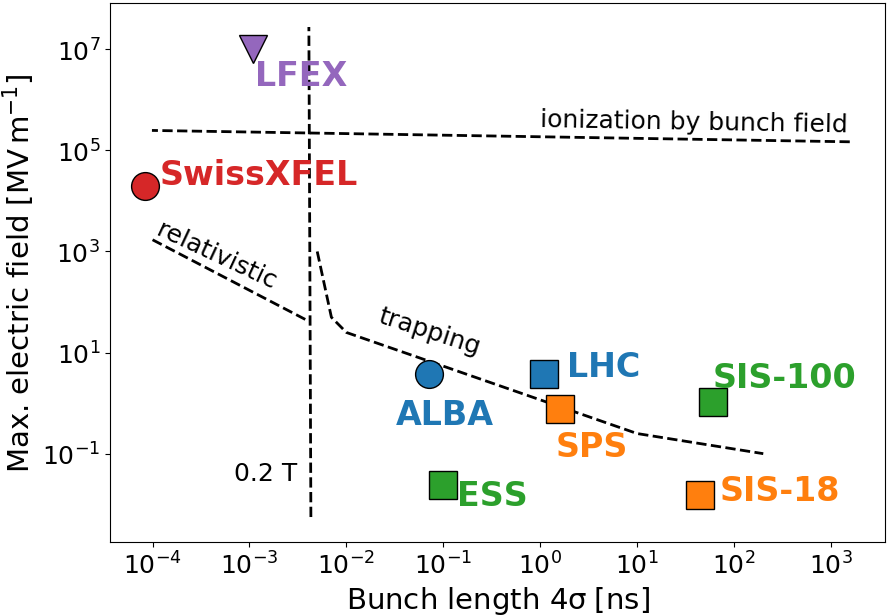}
    \caption[Parameter space]{Amplitude and duration of transient beam electric fields generated in various scientific apparatus.
    Approximate boundaries indicating appearance of phenomena like electron trapping or relativistic regime are shown. The estimates were done for a magnetic guiding field of \SI{0.2}{\tesla}. The dashed vertical line indicates the gyro-period for electrons moving in the \SI{0.2}{\tesla} magnetic field.}
    \label{fig:02:overview}
\end{figure}

Some of the mentioned effects emerge from the influence of the particle beam and depend mainly on the amplitude of the beam electric field and the duration of the corresponding pulse.
An overview of various effects in dependence on these two parameters is shown in \rfig{fig:02:overview} alongside the beam parameters of several hadron accelerators, 
the most powerful laser system up-to-date (LFEX), 3rd generation light source (ALBA) and advanced free electrons laser (SwissXFEL). The peak fields in waist of LFEX laser beam reaches \SI{1d7}{\mega\volt\per\meter}, well beyond the limit of gas ionization by bunch electric field. In comparison, the highest field in an accelerator - a focused SwissXFEL beam - will be reaching about \SI{2d4}{\mega\volt\per\meter}. All hadron machines are currently below \SI{10}{\mega\volt\per\meter} but the bunch lengths are about 3 orders of magnitude larger than those of LFEX and XFEL. \rfig{fig:02:overview} also indicates two effects related to beam space-charge interaction: electrons being trapped by the beam fields exceeding the electric guiding field and electron velocities reaching into the relativistic regime.

Depending on the acquisition system in use, additional distortion of the measured distributions may occur. The electron or ion detection is usually based on amplification of the signal using Multi-Channel Plate (MCP) and either conversion of resulting electrons to light using a phosphor screen followed by a camera or direct conversion to electric signals using anode-strips. Such systems may suffer from non-uniformity of MCP gain, Phosphor deterioration or point-spread function of the optical system positioned after the Phosphor. Recent developments focus on use of hybrid silicon pixel detector instead of MCP/Phosphor or MCP/anode-strip combination and have shown to sidestep some of these issues~\cite{hybrid-pixel-detector}.

The present study focuses on application of electron-based IPMs in high-brightness hadron machines such as the Large Hadron Collider (LHC), the Super Proton Synchrotron (SPS) at CERN or the SIS-100 under construction at FAIR. For these machines the beam-space induced profile distortion becomes a prominent factor that potentially hinders the successful operation of electron-based IPMs. This type of profile distortion as well as possible mitigation strategies are the focus of the present study.

\section{Profile Distortion}
\label{section:profile-distortion}
In the following sections we will focus on profile distortion effects related to the electron movement which is governed by their interaction with the present electromagnetic fields. For the scope of this contribution, we assume perfectly uniform and aligned guiding fields and do not consider gas target and acquisition system related effects which were mentioned in the previous section. The coordinate system used throughout this study is \textit{x}, along the measured profile (horizontal), \textit{y}, towards the acquisition system (vertical) and \textit{z}, along the beam (longitudinal).

For illustration purposes we refer in this section to the data obtained for an example case which has been studied by means of simulations. The corresponding parameters are given in \rtable{table:example-case-parameters} and correspond to a possible LHC flat top configuration (nominal value of bunch charge is about $\rm 1.3\cdot 10^{11}$, but charges above $\rm 2\cdot 10^{11}$ are expected for HL-LHC). The LHC beam case is well studied~\cite{Patecki-Thesis, Sapinski-HB-2014, Vilsmeier-Thesis, IBIC-2017-ML, IPAC-2018-ML, HB-2018-ML}. As another example we will shortly discuss the SIS-100 IPM system for proton beam at flat top.

\begin{table}[htbp]
  \centering
  \caption{Beam and device parameters for the example case which corresponds to a possible LHC flat top configuration.}
  \label{table:example-case-parameters}
  \begin{tabular}{lcc}
      \toprule
         \textbf{Parameter}           & \textbf{Value} \\
      \midrule
         Beam particle type           & Protons \\
         Energy/u                     & \SI{6.5}{\tera\electronvolt} \\
         Bunch population             & \SI{2.1d11}{ppb} \\
         Bunch length ($4\sigma_z$)   & \SI{0.9}{\nano\second} \\
         Bunch width ($1\sigma_x$)    & \SI{270}{\micro\metre} \\
         Bunch height ($1\sigma_y$)   & \SI{360}{\micro\metre} \\
         Electrode distance           & \SI{85}{\milli\metre} \\
         Applied voltage              & \SI{4}{\kilo\volt} \\
         Magnetic field               & \SI{0.2}{\tesla} \\
         Number of sim. electrons     & \num{100000} \\
         Time step size               & \SI{0.3125}{\pico\second} \\
      \bottomrule
  \end{tabular}
\end{table}

\subsection{Effects related to electron movement}

The electron movement is initialized by the ionization process and further influenced by the interaction with the guiding fields and the electromagnetic field of the particle beam (\textit{space-charge interaction}). The relevant effects can be summarized by the following three aspects :

\begin{itemize}
    \item \textit{Ionization momenta} -- The initial momenta of electrons, obtained during the ionization process, clearly influence their further trajectories. Due to the interaction with the magnetic guiding field an initial momentum component along the beam axis results in a transverse displacement of the subsequent gyromotion while a transverse momentum component results in a longitudinal displacement. This effect becomes more pronounced for small guiding field strengths as well as for small beam sizes where the magnitude of the gyromotion increases relative to the beam dimensions. For relativistic beams the electrons are mainly ejected transverse to the beam direction~\cite{voitkiv}.
    \item \textit{Space-charge interaction} -- The interaction of electrons with the electric field of the particle beam may significantly alter their trajectories, resulting in noticeable distortions of measured profiles. This interaction cannot be treated analytically without introducing significant approximations and hence must be studied by the means of numerical simulations. As this effect correlates with the magnitude of the beam electric field, it becomes especially significant for high intensity and high energy beams. However also small magnetic guiding fields may not be sufficient to counteract the beam field interaction and hence provoke an increase in gyroradii.
    \item \textit{Gyromotion} -- The electrons perform a spiral movement under the influence of the external magnetic guiding field. Depending on the magnitude of the gyroradius an electron may experience a significant displacement with respect to its center of gyration. This effect is especially significant for small beam sizes since the magnitude of the displacement increases relative to the beam size.
\end{itemize}

Often the magnetic guiding field is sufficient for suppressing profile distortion. However in extreme cases the space-charge interaction can become so strong, with electric fields of up to a few \si{\mega\volt\per\meter} (typical extraction fields are around \SI{50}{\kilo\volt\per\meter}), that electrons are (a) trapped inside the space-charge region since the beam electric field outweighs the electric guiding field around the center of the bunch, and (b) forced on significantly different trajectories, resulting in a vast increase of gyroradii. Since the above effects depend on the various parameters of the particle beam it is often not obvious to determine an appropriate strength for the magnetic guiding field which allows to sufficiently suppress the gyroradius increase.

Having in mind the above-mentioned effects we may subdivide the IPM volume into two regions:
\begin{itemize}
    \item \textit{Space-charge region} -- This is the region in the vicinity of the beam where the interaction with the beam's electromagnetic field has a major influence. The gyro-velocity of electrons are subject to a perpetual oscillation induced by the \ExB drift along the beam and the magnitude of this oscillation is altered when the beam center approaches and recedes from the positions of the electrons. Because of the complex shape of the electromagnetic fields the electron motion exhibits other electromagnetic drifts as well, for example \textit{polarization drift} due to the time dependence of the beam electric field~\cite{electromagnetic-drifts}. For estimating the dimensions of this region one can consider the \ExB drift velocity as an indicator for space-charge action. The region in which this \ExB drift velocity is more than 1\% of the undisturbed gyro-velocity, is of the order of a few millimeters.
    \item \textit{Detector region} -- This is the region close to the acquisition system where the beam fields diminish to a negligible magnitude. This means the electron movement are solely subject to the electric and magnetic guiding fields and hence the electrons perform a pure gyromotion planar to the detector while being accelerated towards it. The characteristics of this gyromotion (gyroradius and gyrocenter position) depend on the previous beam space-charge interaction.
\end{itemize}

\begin{figure}[htbp]
  \centering
  \includegraphics[width=8.6cm, keepaspectratio]{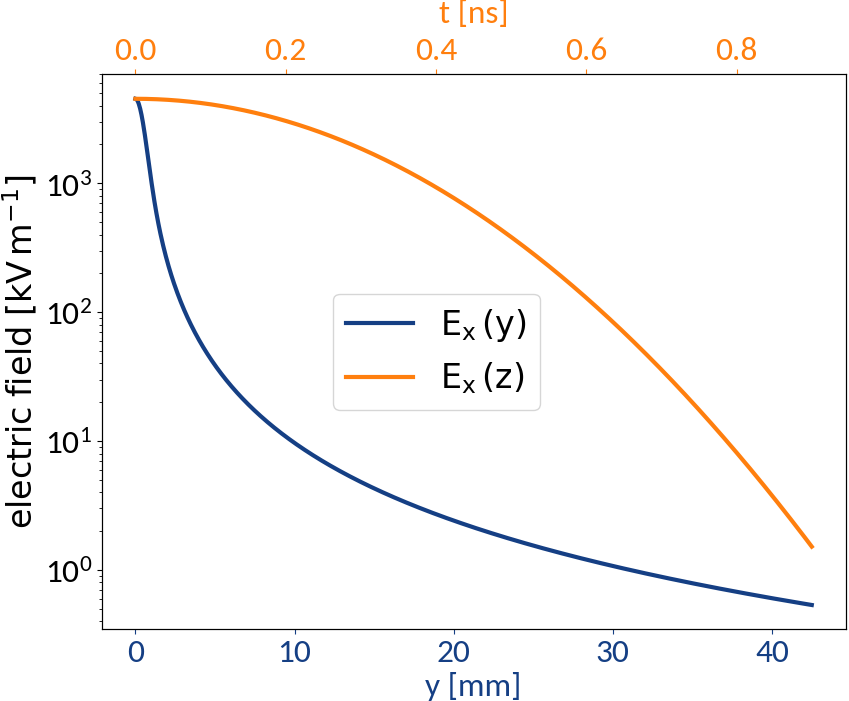}
  \caption{Evolution of the transverse bunch electric field $E_x$ at $x = \sigma_x$ in direction towards the detector (y) and during beam passage (z, t), for the parameters in \rtable{table:example-case-parameters}. The x-position is fixed to $1\,\sigma_x$ and y- and z-position are fixed to zero if not varied.}
  \label{fig:electric-field-long-distance}
\end{figure}

Note that as the bunch recedes from the electrons' positions the space-charge region will shrink in time. While this region can be quite large during the bunch center passing, the transverse electric field quickly diminishes according to the bunch's line density. Therefore also electrons that are ionized near the bunch's center will eventually end up performing a pure gyromotion as in the detector region.
\rfig{fig:electric-field-long-distance} shows the bunch electric field as it decreases towards the detector ($y$) and along the beam axis ($z$) and hence illustrates the separation in space-charge and detector region. While the two dependencies are shown separately, their collective effect results in a much stronger decrease as the bunch recedes from the electrons' positions and hence the effective space-charge region is smaller than indicated by the single dependencies. This is because the beam velocity is significantly larger than the electrons' velocities along the beam axis and hence the spatial dependence ($z$) can be thought of as a time-dependence ($t$).

Considering an electron's \textit{initial} position during ionization and \textit{final} position during detection, $x_0$ and $x_f$ respectively, the overall displacement $\Delta x = x_f - x_0$ can be ascribed to three different effects:

\begin{enumerate}
    \item $\Delta x_1$ -- Displacement of the electron's gyro-center $x_0^c$ with respect to the initial position, due to its initial velocity along the beam: $\Delta x_{1} = x_0^c - x_0$. Electrons that are ejected transverse to the longitudinal beam axis, in horizontal direction, merely suffer from a displacement of their gyrocenter along that axis, i.e. not affecting their horizontal position which determines the transverse profile. If they are ejected along the beam however there will be a corresponding displacement transverse to it. This type of displacement is typically around a few tens of microns and can be assessed with differential ionization cross sections which usually depend on the projectiles charge state and energy. In relativistic cases, transverse ejection dominates and so the resulting displacement is mainly along the beam~\cite{voitkiv}.
    \item $\Delta x_2$ -- Displacement of the space-charge altered gyrocenter $x_s^c$ with respect to the initial gyrocenter: $\Delta x_{2} = x_s^c - x_0^c$. This shift is found to be induced by the magnetic field of the beam. In addition the longitudinal electric field of the beam induces an \ExB drift in horizontal direction which adds up as an additional displacement of the gyrocenter. However for relativistic beams the exposure to the longitudinal field is short and hence this drift tends to be negligible. The net effect of this type of displacement is around a few microns even for high beam currents.
    \item $\Delta x_3$ -- Displacement of the final position with respect to the stable gyrocenter due to the nature of the gyromotion: $\Delta x_{3} = x_f - x_s^c$. The electrons perform a gyromotion in the plane above the detector and in case this motion spans over multiple elements of the detecting system, the electrons might be detected on any of them. If the moment of detection is not correlated to the phase of the gyromotion then the resulting displacement is random. Since this type of displacement is proportional to the resulting gyroradii it depends on the magnitude of the preceding space-charge interaction and can be around some hundreds of microns up to a few millimeter.
\end{enumerate}

\begin{figure}[htbp]
  \centering
  \includegraphics[width=17.8cm, keepaspectratio]{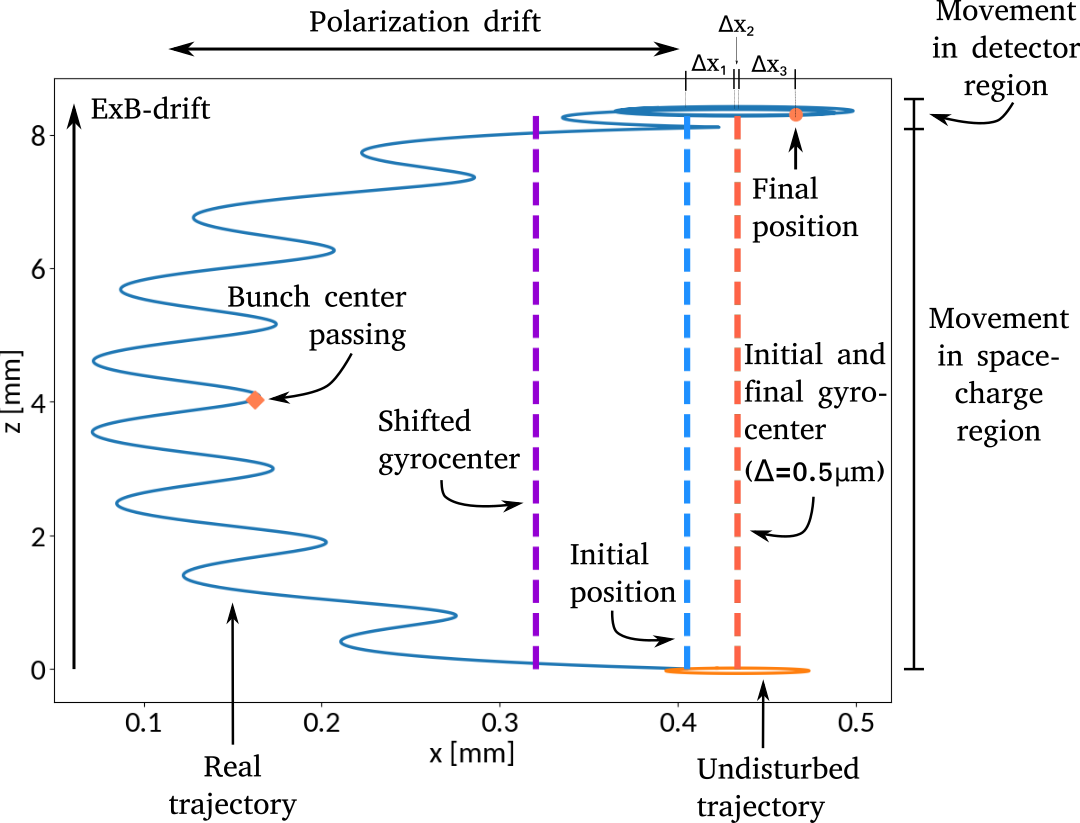}
  \caption{Example trajectory obtained for $\bm{r_0} = (1.5\sigma_x, 0, 0), \bm{v_0} = 10^{6}\cdot (-1, 0, 1) \si{\meter\per\second}$ and generated $2\sigma_z$ before the bunch center (beam parameters corresponding~\rtable{table:example-case-parameters}). The term "undisturbed trajectory" refers to the case if there were no space-charge interaction, i.e. only pure gyromotion. Because the initial and final gyroradius is about \SI{100}{\micro\meter} the corresponding part of the trajectory appears as a flat ellipse due to large scale of the z-drift. The "shifted gyrocenter" is induced by the transverse electric field of the beam and acts as a starting point for the polarization drift; this shift is described by equation~\req{eq:shifted-gyrocenter}. "Bunch center passing" indicates the moment when the electron and the bunch center are aligned with respect to z-position. The "initial" and "final gyrocenter" are separated by a distance of \SI{0.5}{\micro\meter}. The labels ``Movement in detector / space-charge region'' refer to the period that is spent in each of the regions, not the spatial domain, as the vertical axis indicates movement along the beam axis. There is no transverse \ExB drift because the simulation was configured with $E_z = 0$.}
  \label{fig:example-trajectory}
\end{figure}

\rfig{fig:example-trajectory} shows an example trajectory which exhibits the different electromagnetic drifts, the separation in space-charge and detector region, as well as the different kind of displacements $\Delta x_{\{1, 2, 3\}}$ mentioned above. The increase in gyroradius is also clearly visible. The maximum polarization drift distance, corresponding to the parameters in \rtable{table:example-case-parameters}, is estimated to be \SI{852}{\micro\meter} as further discussed in appendix~\ref{section:estimation-of-polarization-drift-distance}.

These effects only consider displacements in horizontal direction, along the measured profile. In fact the electrons experience an additional significant displacement in longitudinal direction, along the beam, due to the \ExB drift  induced by the transverse electric field of the beam $E_x$. However such drifts can be neglected in case the physical situation is similar along the beam axis (i.e. uniform guiding fields, similar gas pressure). In typical setups the size of the readout system is much smaller than the corresponding area of the surrounding IPM field box and it is centered therein. Hence these conditions are satisfied.

\begin{figure}[htbp]
  \centering
  \includegraphics[width=8.6cm, keepaspectratio]{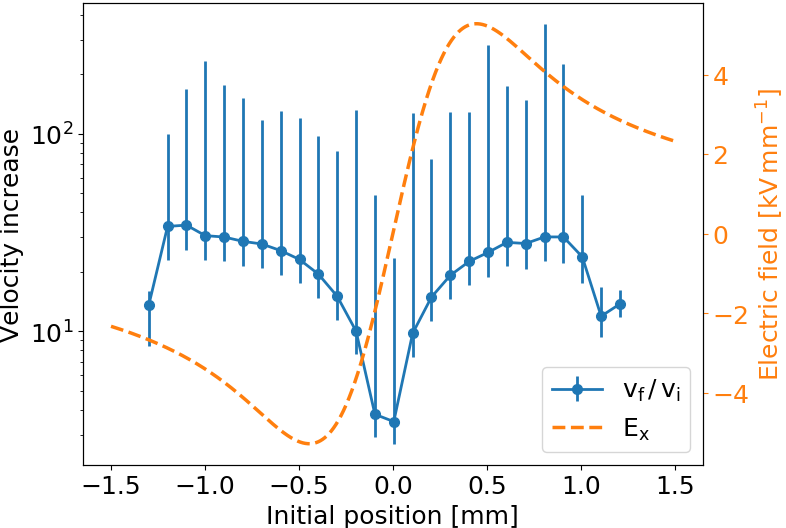}
  \caption{Velocity increase, given as the ratio of velocity at detection and velocity at ionization, plotted against the initial horizontal position of electrons. Dots denote average values while vertical bars indicate 1 standard deviation. The transverse electric field in x-direction, measured at $y = z = 0$ at the center of the bunch, is overlaid. Beam parameters corresponding to \rtable{table:example-case-parameters}.}
  \label{fig:velocity-increase-vs-initial-position}
\end{figure}

\begin{figure}[htbp]
  \centering
  \includegraphics[width=8.6cm, keepaspectratio]{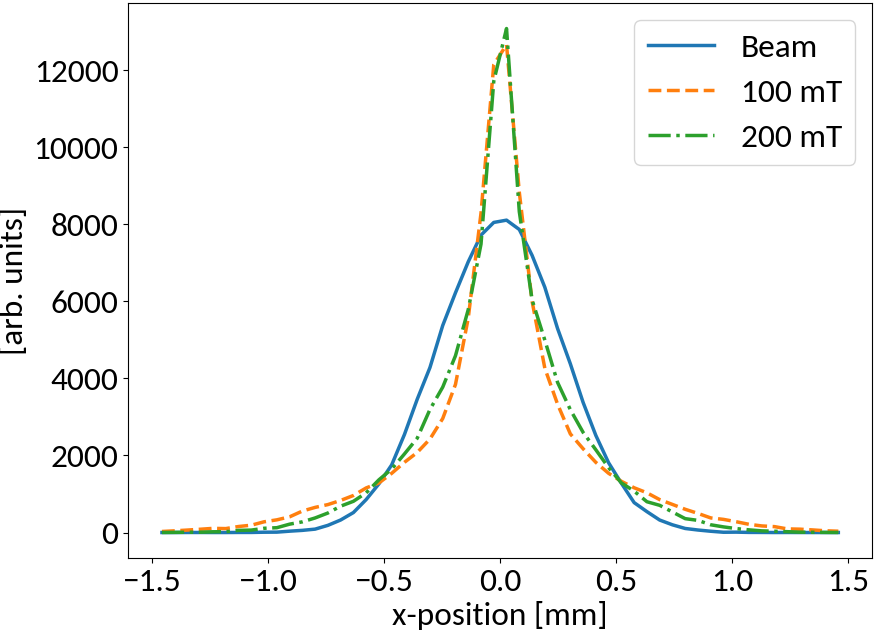}
  \caption{Simulated example profiles for beam parameters corresponding to \rtable{table:example-case-parameters} using magnetic guiding field strengths of \SI{100}{\milli\tesla} and \SI{200}{\milli\tesla} respectively. The beam standard deviation is \SI{270}{\micro\meter} and the resulting standard deviations of the simulated profiles are \SI{366}{\micro\meter} and \SI{305}{\micro\meter} respectively.}
  \label{fig:example-profiles}
\end{figure}

\rfig{fig:velocity-increase-vs-initial-position} shows the velocity increase of electrons which is induced by the space-charge interaction and consequentially leads to a deformation of measured profiles as shown by \rfig{fig:example-profiles}. The mean of initial velocities is \SI{9.94d5}{\meter\per\second} which corresponds to a gyroradius of \SI{28.3}{\micro\meter}.
The total displacement of an electron is compound of the three above-mentioned stages: $\Delta x = \Delta x_1 + \Delta x_2 + \Delta x_3$. Regarding the transverse \ExB drift due to non-zero longitudinal field component $E_z$ one can consider the fact that the transverse field is scaled with relativistic factor of $\gamma \approx \num{7000}$. Since the \ExB drift velocity is proportional to the electric field strength and the longitudinal drifts typically are of the order of a few millimeter the expected \ExB drift distance transverse to the beam axis is expected to be at least $\gamma^{-1}$ times smaller, resulting in transverse \ExB drift distances in the sub micrometer regime.

\subsection{Description for uniform beam distributions}\label{sec:analytic-considerations-for-uniform-beams}

Electron trajectories are mainly influenced by the transverse electric field of the beam and the magnetic guiding field. This interaction alters the trajectories in a complex way which cannot be predicted analytically for any realistic beam distribution without further assumptions. 

\begin{figure}[htbp]
    \centering
    \includegraphics[width=17.8cm]{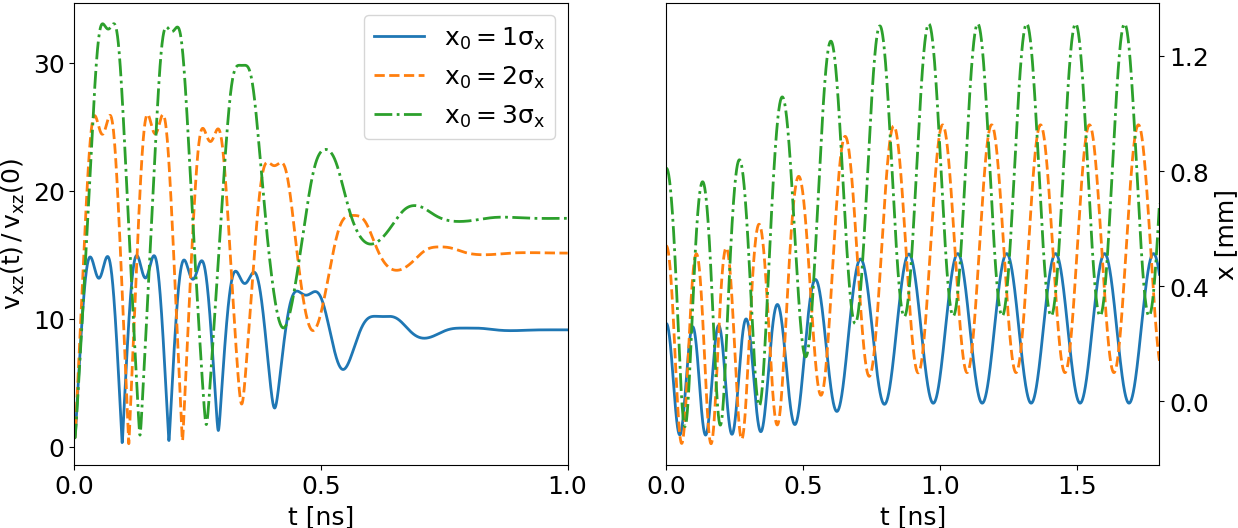}
    \caption[Gyro-velocity evolution for different $x_0$]{Left: Gyro-velocity evolution for different starting positions $\bm{r_0} = (x_0, 0, 0)$. Initial velocity is $\bm{v_0} = 10^{6}\sqrt{2}^{-1} \cdot (1, 1, -1) \si{\meter\per\second}$ and electrons are generated $\sigma_z / 2$ before the bunch center (beam parameters corresponding~\rtable{table:example-case-parameters}). Right: The corresponding x-position evolution which turns into pure gyromotion for $t\gtrapprox \SI{1}{\nano\second}$.}
    \label{fig:02:velocity-trajectories-various-x0}
\end{figure}

In order to get an analytic estimate of the interaction we assume a simplified case in which the longitudinal electric field is zero and the transverse electric field is approximated by a linearly increasing field $E_x(x) = E\cdot x$ where  $E < 0$ (this corresponds to the electric field inside a uniformly charged cylinder bunch of positive ions).
For a Gaussian bunch shape this approximation is also well applicable at the center of the bunch as can be seen from \rfig{fig:velocity-increase-vs-initial-position} (neglecting the y- and z-dependence of the field). We also consider no bunch magnetic field since it is small compared to the magnetic guiding field $B$ which is acting in y-direction. The electric guiding field $E_g$ is acting in y-direction as well and hence has no influence on the motion in the xz-plane.
In this simplified scenario the particle motion in the xz-plane is obtained as:

\begin{flalign}
\label{eq:uniform-bunch-motion-x}
 x(t)       &= \frac{v_{x0}}{\Omega}\sin(\Omega t) + \frac{\bar{q}}{\Omega^2}(Bv_{z0} - Ex_0)\cos(\Omega t)  + \left(1 + \frac{\bar{q}E}{\Omega^2}\right)x_0 - \frac{\bar{q}B}{\Omega^2}v_{z0} 
\end{flalign}

\begin{flalign}
\label{eq:uniform-bunch-motion-z}
 z(t)       &= -\frac{\bar{q}B}{\Omega^2}v_{x0}\cos(\Omega t) +\frac{\bar{q}^2B}{\Omega^3}(Bv_{z0} - Ex_0)\sin(\Omega t) 
+ \left[ \left( \frac{E}{B} + \frac{\bar{q}E^2}{\Omega^2B} \right)x_0 - \frac{\bar{q}E}{\Omega^2}v_{z0} \right]t \\ \nonumber
            &\; + z_0 + \frac{\bar{q}B}{\Omega^2}v_{x0}
\end{flalign}

\begin{flalign}
\label{eq:uniform-bunch-motion-vx}
 \dot{x}(t) &= v_{x0}\cos(\Omega t) - \frac{\bar{q}}{\Omega}(Bv_{z0} - Ex_0)\sin(\Omega t) 
\end{flalign}

\begin{flalign}
\label{eq:uniform-bunch-motion-vz}
 \dot{z}(t) &= \frac{\bar{q}B}{\Omega}v_{x0}\sin(\Omega t) + \frac{\bar{q}^2B}{\Omega^2}(Bv_{z0} - Ex_0)\cos(\Omega t) 
             + \left( \frac{E}{B} + \frac{\bar{q}E^2}{\Omega^2B} \right)x_0 - \frac{\bar{q}E}{\Omega^2}v_{z0} &&
\end{flalign}

where we have used the following abbreviations ($q$ is the elementary charge and $m$ is the electron mass):
\begin{subequations}\label{eq:definitions}
 \begin{align}
  \bar{q} &\equiv \frac{q}{m} \\
  \Omega^2 &\equiv \bar{q}^2B^2 - \bar{q}E > 0\;,\;\; \Omega = \sqrt{\Omega^2}
 \end{align}
\end{subequations}
\req{eq:uniform-bunch-motion-x} shows that the particle performs an oscillating movement in x-direction while \req{eq:uniform-bunch-motion-z} shows a similar oscillation as well as an \ExB drift in z-direction.
The absolute value of the gyro-velocity is given by $ \lvert\bm{v_{xz}(t)}\rvert = \sqrt{\dot{x}(t)^2 + \dot{z}(t)^2}$. From \req{eq:uniform-bunch-motion-vx} and \req{eq:uniform-bunch-motion-vz} we can see that this velocity expression contains terms $\sin^2(\Omega t)$, $\cos^2(\Omega t)$ and $\sin(\Omega t)\cdot\cos(\Omega t)$ which are $\pi$-periodic as well as terms $\sin(\Omega t)$ and $\cos(\Omega t)$ which are $2\pi$-periodic. \rfig{fig:02:velocity-trajectories-various-x0} shows the gyro-velocity evolution for three example particles subject to the realistic electric field of a Gaussian bunch. The expected twofold periodicity is clearly visible while an additional damping effect occurs due to the longitudinal field dependence when the beam recedes from the particles' positions. The minima and maxima of the velocities become apparent when considering the two-dimensional trajectories $(x(t), z(t))$.
At one turn-around point the velocities from \ExB drift and the oscillating movement are aligned and add up, resulting in a large velocity while at the other turn-around point they are opposite resulting in a smaller velocity (or even a backward drift). 
Such behavior can as well be observed from simulations as described below.
\rfig{fig:02:velocity-trajectories-various-x0} shows the velocities of various particles for different starting points $x_0$ subject to the realistic field of a Gaussian charge distribution. One can observe a variation of the amplitude modulation as well as different periods of the oscillation depending on the starting point. This is due to the non-linear shape of the Gaussian-bunch electric field for positions farther away from the bunch center (compare the electric field plot in \rfig{fig:velocity-increase-vs-initial-position}).

Analyzing \req{eq:uniform-bunch-motion-x} to \req{eq:uniform-bunch-motion-vz} we can infer various parameters of the resulting electron motion.

\subsubsection{Shift of gyration center}
From \req{eq:uniform-bunch-motion-x} one can read the center of oscillation
\begin{equation}\label{eq:shifted-gyrocenter}
 x_c = \left( 1 + \frac{\bar{q}E}{\Omega^2} \right)x_0 - \frac{\bar{q}B}{\Omega^2}v_{z0}
\end{equation}
which contains an additional term corresponding to the transverse bunch electric field.
This shift however is compensated due to the time-dependence of the electric field and the corresponding polarization drift as shown in Appendix \ref{section:gyrocenter-shift-compensation}.
The same effect can be observed from simulations as well.
This results in the initial and final gyrocenter to be aligned with respect to each other as shown in \rfig{fig:example-trajectory}. Note that a displacement of the gyrocenter with respect to the initial position $x_0$ however occurs due to a non-zero initial velocity component along the beam $v_{z0}$. This shift is not compensated for by any other effects.

\subsubsection{Displacement due to gyromotion}
\begin{figure}[htbp]
    \centering
    \includegraphics[width=8.6cm]{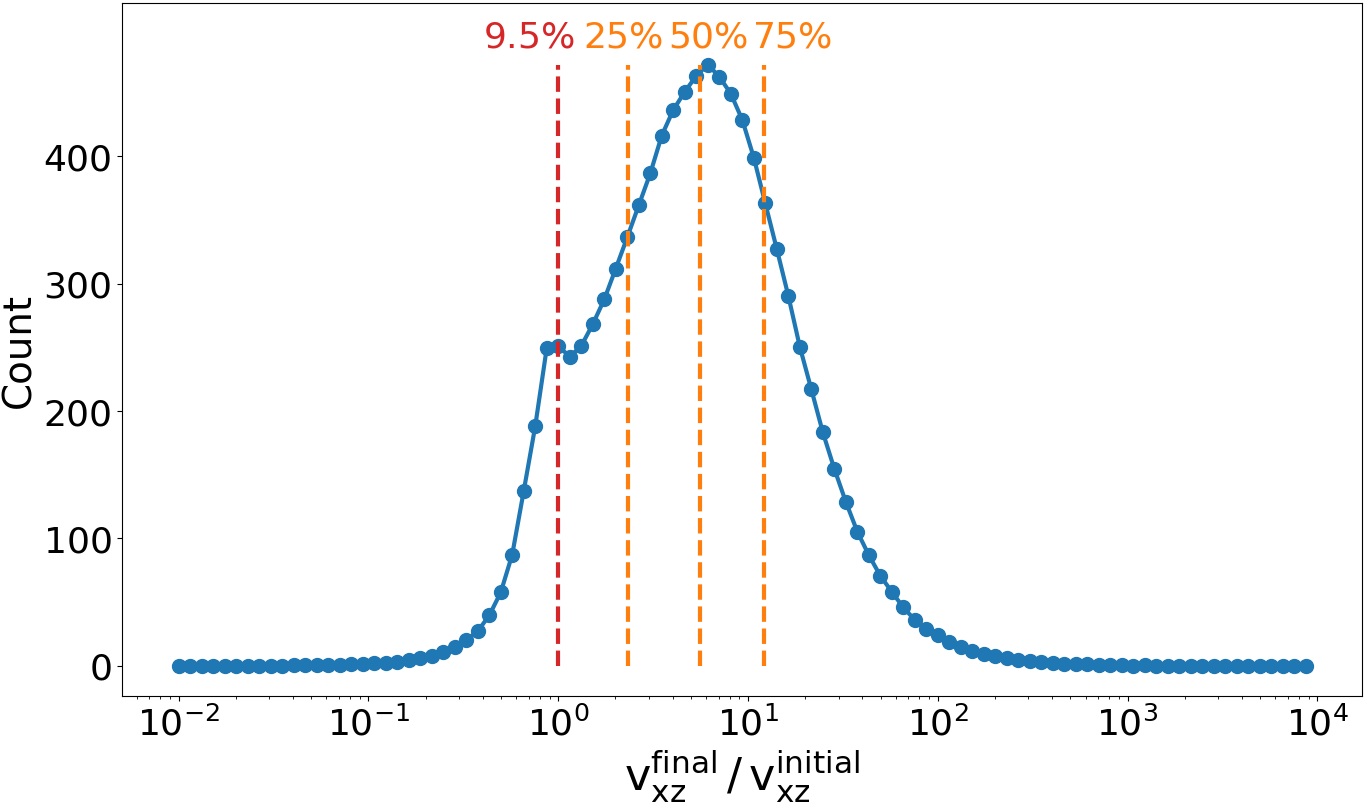}
    \caption{Distribution of gyro-velocity ratios, final over initial velocity, due to space-charge interaction (parameters corresponding to \rtable{table:example-case-parameters}).}
    \label{fig:02:velocity-increase}
\end{figure}

In the detector region, without the influence of the beam electric field, the electrons perform a pure gyromotion with gyroradius proportional to the planar velocity.
Because of the velocity oscillations in the space-charge region, described by \req{eq:uniform-bunch-motion-vx} and \req{eq:uniform-bunch-motion-vz} as well as \rfig{fig:02:velocity-trajectories-various-x0}, the electrons are likely to end up with an increased velocity in the detector region and consequentially with an increased gyroradius.
\rfig{fig:02:velocity-increase} shows the distribution of gyro-velocity change. Around 90\% of electrons end up with an increased gyro-velocity due to space-charge interaction and hence end up with increased gyroradii.

The resulting gyromotion implies possible displacements to positions different from the center of the gyromotion. This effect is discussed in detail in~\cite{Sapinski-HB-2014, Vilsmeier-Thesis}.

\subsubsection{Time-of-flight}
In order to obtain an estimate for the final gyroradius of electrons we need to derive an estimate for their time-of-flight until they reach away from the influence of disturbing beam fields.
Since the above analytic considerations are valid only inside the bunch, we set the space-charge region as the region inside the bunch and the detector region as the region outside the bunch. This assumption neglects modifications of the particles' velocities outside the bunch while in reality they may remain affected by the bunch electric field at these positions as the space-charge region may extend farther out.
A more complex description would be to treat the decreasing part of the field outside the bunch by another linear field model, similarly to the one inside, and to reuse the above derived equations for that case by applying appropriate coordinate transformations. However since this approach is significantly more complex we stick with the former approach as it turns out to be sufficient as indicated by the following simulation results.

For estimating the time-of-flights one cannot simply rely on the time the bunch needs to recede from the electron but one needs to consider the electron leaving the bunch volume (i.e. the space-charge region) in the transverse direction as well. In order to obtain a more accurate estimate for this time-of-flight in the vertical direction we need to take the bunch electric field and the electric guiding field into account. We do so by simplifying the longitudinal time dependence by an exponential dependency instead of the Gaussian dependency, in order to relax the explicit time dependence. The spatial dependence is linear as before. The corresponding equation of motion is:
\begin{equation}
    \ddot{y}(t) = \bar{q}E_g - \bar{q}E y \exp\left(-\frac{\lvert t\rvert}{\sigma_z}\right)
\end{equation}

where $E_g$ is the electric guiding field strength.

For a detailed derivation of the resulting motion $(y(t), \dot{y}(t))$ consider Appendix~\ref{section:tof-for-uniform-shape}.

In order to estimate the time-of-flights we need to find the roots of the functions $f_1(t) = \sqrt{x(t)^2 + y(t)^2} - R$ where $R$ is the radius of the cylinder bunch and $x(t)$ is given by \req{eq:uniform-bunch-motion-x} and $f_2(t) = z(t) - (t - 4\sigma_z)\beta c$ where $\beta = v/c$, with $c$ the speed of light; whatever value of $t$ is smaller determines the moment when the electron leaves the bunch. The final parameters are then obtained by plugging this value back into the analytic velocity equations \req{eq:uniform-bunch-motion-vx} and \req{eq:uniform-bunch-motion-vz}. For more details on the computation consider Appendix~\ref{sec:appendix:methods}.

\begin{figure}[htbp]
    \centering
    \includegraphics[width=15.8cm]{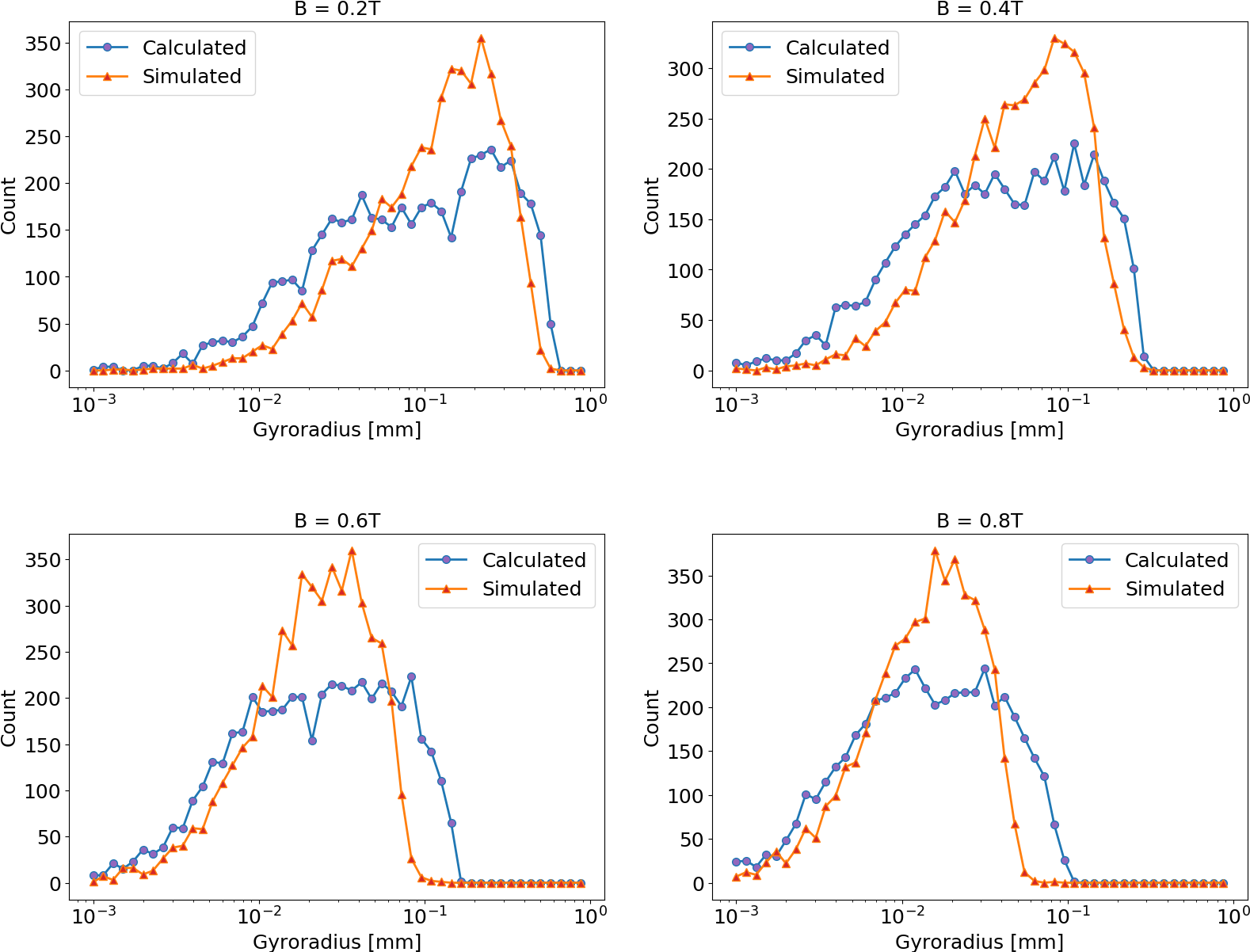}
    \caption{Comparison of gyroradius distributions calculated from analytic equations with distributions obtained from complete simulations with realistic Gaussian bunch fields. Simulation parameters corresponding to \rtable{table:example-case-parameters} and the magnetic guiding field is varied.}
    \label{fig:02:gyroradius-distributions-analytical-compared-with-simulation}
\end{figure}

\rfig{fig:02:gyroradius-distributions-analytical-compared-with-simulation} shows the gyroradius distributions computed the analytic estimation for different magnetic field strengths, compared with the results obtained from complete simulations including all effects by using realistic Gaussian bunch fields. The beam and device parameters correspond to \rtable{table:example-case-parameters}. The simulations were performed using the Virtual-IPM simulation tool~\cite{virtual-ipm}. The estimated distributions show agreement in the shape as well as the mean and variance. The additional tails of the calculated distributions emerge as a result of the electric field strength being independent of time in \req{eq:uniform-bunch-motion-x} through \req{eq:uniform-bunch-motion-vz}. Hence for electrons which are ionized before the bunch center the electric field is underestimated and it is overestimated for electrons which are ionized after the bunch center, leading to a corresponding under- and overestimation of the resulting gyroradii which manifests itself in form of the tails of the distribution. The overall agreement of analytic prediction and simulation suggests the applicability of the derived equations. These equations, even though obtained for simplified bunch shape, can be used to further study the phenomenon of profile distortion as shown in the following section.

\subsection{Magnetic field tuning and scaling for realistic beams}

Since IPMs find application in different circumstances including a variety of beam parameters it is useful to establish a relation between the beam parameters and the magnetic field required to suppress profile distortion. Such a relation allows for simple checks when designing new devices.
Different criteria for a sufficient field strength are possible such as absolute thresholds ensuring that most of the electrons' gyroradii are below an acceptable threshold (e.g. requiring a certain fraction to be below some percentage of the bunch width or to be smaller than the detector resolution) or relative thresholds that ensure that the increase of measured profile standard deviation remains below some percent threshold. A previous study considered the gyroradius increase as a criterion~\cite{Patecki-Thesis} but since the relation between a particular gyroradius distribution and the resulting measured beam profile is not obvious we will focus on the standard deviation increase as a more direct measure for the further investigations. Note that due to the nature of profile distortion, which results in enlarged tails but also a heavier peak, already small values of standard deviation increase correspond to strongly non-Gaussian profile shapes as depicted in \rfig{fig:example-profiles}.
We establish a fit of the three most significant beam parameters relevant for profile distortion, the transverse and longitudinal bunch size and the bunch charge:

\begin{equation}
\label{eq:fit-model}
    B_{\textrm{min, 1\%}} = \frac{N^a}{\sigma_z^b\sigma_t^c} \cdot d + \frac{f}{\sigma_t^e}
\end{equation}

where $a,b,c,d,e,f$ are fit parameters, $\sigma_t$ is the standard deviation of the (symmetric) transverse beam distribution to be measured in units units of millimeter, $\sigma_z$ is the standard deviation of the bunch line density in units of nanosecond and $N$ is the number of charges per bunch in units of \num{1d12}. $B_{\textrm{min, 1\%}}$ is the minimum magnetic field required to obtain $\sigma_t^{\textrm{measured}}$ with at least 1\% accuracy with respect to $\sigma_t^{\textrm{beam}}$. The first term in \req{eq:fit-model} corresponds to the space-charge interaction while the second term is attributed to the effect of the initial velocity distribution which, as a constant effect, has a relatively increasing effect on smaller beam profiles; hence the dependence on $\sigma_t$. The accuracy threshold was chosen to be a rather challenging 1\% because this would be useful in investigation of emittance blow-up in modern colliders~\cite{Kuhn2014}. In a subsequent fit this threshold is converted to a parameter.

For verifying the relationship and to infer the parameters we use the analytic considerations from the previous section as well as simulations with three different bunch shapes: Gaussian, Uniform and Parabolic Ellipsoid. For simplicity the bunch shapes are considered to be rotational symmetric around the beam axis ($\sigma_t \equiv \sigma_x = \sigma_y$). The charge density distributions of these shapes are given by:

\begin{equation}
\begin{aligned}
\textrm{\textbf{Gaussian}} \hspace{0.5cm} & \propto \exp\left(-\frac{x^2 + y^2}{2\sigma_t^2} - \frac{z^2}{2\sigma_z^2}\right) \\
\textrm{\textbf{Uniform}} \hspace{0.5cm} &
\begin{cases}
\propto \exp\left(-\frac{z^2}{2\sigma_z^2}\right) &, \textrm{if } x^2 + y^2 \leq \sigma_t^2, \textrm{with radius } \sigma_t \\
\equiv 0 &, \textrm{otherwise}
\end{cases} \\
\textrm{\textbf{Parabolic Ellipsoid}} \hspace{0.5cm} & 
\begin{cases}
\propto 1 - \frac{x^2 + y^2}{\sigma_t^2} - \frac{z^2}{\sigma_z^2} & , \textrm{if } \frac{x^2 + y^2}{\sigma_t^2} + \frac{z^2}{\sigma_z^2} \leq 1, \textrm{with semi-axes } \sigma_z, \sigma_t \\
\equiv 0 & , \textrm{otherwise}
\end{cases}
\end{aligned}
\end{equation}

We only consider relativistic beams since for low energy beams the ionization cross sections take on substantially different shapes~\cite{rudd-et-al}, the generated electrons are exposed to the beam electric field for a longer period due to the reduced beam velocity and thus also effects from the longitudinal electric field start to play a role. We used relativistic $\beta \geq 0.99$ when generating the fit data.

The relationship \req{eq:fit-model} is fitted with simulation data obtained with the Virtual-IPM simulation tool, for the parameters given in \rtable{table:fit-region}. These parameters correspond to the Gaussian bunch shape. For the other bunch shapes the parameters are scaled where appropriate in order to minimize the squared difference between initial beam distributions; for more details consider Appendix~\ref{section:bunch-shape-scaling-factors}. The resulting scaling factors are given in \rtable{table:fit-region} as well.
\rtable{table:fit-results} shows the fit parameters obtained for the various bunch shapes as well as the quality of fit. They show agreement among each other and also with the results obtained for the analytic derivations. Applying the formula to the example case in \rtable{table:example-case-parameters} we obtain a minimum required field of \SI{536}{\milli\tesla} for suppressing the increase in standard deviation below \SI{1}{\percent}. A previous study, supported by a different simulation tool, estimated the minimum magnetic field required for distortion-free measurements to be \SI{1}{\tesla}~\cite{Sapinski-HB-2014}. Using these two magnetic field strengths in our simulations we obtain an increase in standard deviation of \SI{0.6}{\percent} and \SI{0.08}{\percent} respectively which is in conformity with \req{eq:fit-model} and the results of the previous study. Applying the same formula to a set of beam parameters corresponding to the planned IPM on SIS-100 proton beam (\SI{29.9}{\giga\electronvolt} energy, \num{2d13} charges, $\sigma_t = \SI{2.17}{\milli\meter}$, $\sigma_z = \SI{15}{\nano\second}$), the minimum required field is estimated to \SI{76}{\milli\tesla} while the current design foresees a field of \SI{84}{\milli\tesla}~\cite{sis-100-ipm}.

\begin{table}[htbp]
 \centering
 \caption{Beam parameter ranges for simulation of the different bunch shapes. Scaling factors for minimizing the squared difference between the initial one-dimensional profiles of the different shapes are indicated as well. Each parameter follows a logarithmic distribution over 20 points resulting in a total of \num{8000} data points.}
 \label{table:fit-region}
 \begin{tabular}{lccc}
   \toprule
     \textbf{Bunch} & \textbf{-width $\sigma_t$} & \textbf{-length $\sigma_z$} & \textbf{-charge $N$} \\
   \midrule
     Gaussian & \SIrange{0.2}{20}{\milli\meter} & \SIrange{0.3}{300}{\nano\second} & \numrange{1d9}{1d13} \\
     Uniform & $\times \; 1.76$ & $\times \; 1$ & $\times \; 1$ \\
     Parabolic Ellipsoid & $\times \; 2.44$ & $\times \; 2.44$ & $\times \; 1$ \\
   \bottomrule
 \end{tabular}
\end{table}

A previous derivation of minimum required magnetic field strength that was done for related circumstances, however considering the gyroradius increase as an intermediate quantity instead of the final profile standard deviation, shows a similar dependence on the beam parameters~\cite{Patecki-Thesis}.

\begin{table}[htbp]
  \centering
  \caption{Fit parameters for \req{eq:fit-model} for different bunch shapes as well as for the analytic considerations of section \ref{sec:analytic-considerations-for-uniform-beams}. The quality of fit in form of mean squared error (MSE) is given as well.}
  \label{table:fit-results}
  \begin{tabular}{lcccccc|c}
      \toprule
         & \textbf{a} & \textbf{b} & \textbf{c} & \textbf{d} & \textbf{e} & \textbf{f} & \textbf{MSE [\si{\tesla\squared}]} \\
      \midrule
         Analytic  & 0.555 & 0.555 & 0.998 & 0.163 & 1.007 & 0.032 & \num{3.65d-04} \\
         Uniform   & 0.609 & 0.605 & 1.013 & 0.110 & 0.986 & 0.049 & \num{8.57d-04} \\
         Gaussian  & 0.613 & 0.590 & 1.082 & 0.105 & 0.967 & 0.038 & \num{6.35d-04} \\
         Parabolic & 0.623 & 0.595 & 1.027 & 0.103 & 0.974 & 0.046 & \num{8.76d-04} \\
      \bottomrule
  \end{tabular}
\end{table}

In order to assess the effect of a chosen threshold for standard deviation increase $\tau = \sigma_m\sigma_t^{-1} - 1$, with measured (simulated) standard devation $\sigma_m$, on the minimum required magnetic field we include this parameter as well in a subsequent fit of the form:

\begin{equation}
\label{eq:fit-model-with-threshold}
    B_{\textrm{min}} = \frac{N^a}{\sigma_z^b\sigma_t^c\tau^g} \cdot d + \frac{f}{\sigma_t^e\tau^h}
\end{equation}

The results are similar to the ones obtained before and are shown in \rtable{table:fit-results-with-threshold}.

\begin{table}[htbp]
  \centering
  \caption{Fit parameters for equation \req{eq:fit-model-with-threshold} for the Gaussian bunch shape. The quality of fit in form of mean squared error (MSE) is given as well.}
  \label{table:fit-results-with-threshold}
  \begin{tabular}{cccccccc|c}
      \toprule
         \textbf{a} & \textbf{b} & \textbf{c} & \textbf{d} & \textbf{e} & \textbf{f} & \textbf{g} & \textbf{h} & \textbf{MSE [\si{\tesla\squared}]} \\
      \midrule
         0.592 & 0.559 & 1.052 & 0.020 & 0.938 & 0.005 & 0.384 & 0.479 & \num{3.79d-04} \\
      \bottomrule
  \end{tabular}
\end{table}

Another way to include the threshold in the formula is to model the remaining fit parameters as dependent on the threshold:

\begin{equation}
\label{eq:fit-model-with-threshold-dependency}
    B_{\textrm{min}} = \frac{N^{a(\tau)}}{\sigma_z^{b(\tau)}\sigma_t^{c(\tau)}} \cdot d(\tau) + \frac{f(\tau)}{\sigma_t^{e(\tau)}}
\end{equation}

This formulation captures potential variations in the beam parameter dependence for varying thresholds $\tau$. The results however, as shown in \rtable{table:fit-results-with-threshold-dependency}, confirm that the effect of varying threshold is mainly a scaling factor. These results are in agreement with the $\tau^f$ scaling shown from \req{eq:fit-model-with-threshold} and \rtable{table:fit-results-with-threshold}.

\begin{table}[htbp]
  \centering
  \caption{Fit parameters for equation \req{eq:fit-model-with-threshold-dependency} for the Gaussian bunch shape. The quality of fit in form of mean squared error (MSE) is given as well. For $\tau = \SI{1}{\percent}$ the results slightly deviate from the ones in \rtable{table:fit-results} because the underlying simulation data was regenerated.}
  \label{table:fit-results-with-threshold-dependency}
  \begin{tabular}{lcccccc|c}
      \toprule
         Threshold [\%] & \textbf{a} & \textbf{b} & \textbf{c} & \textbf{d} & \textbf{e} & \textbf{f} & \textbf{MSE [\si{\tesla\squared}]} \\
      \midrule
1  &  0.626 &  0.584 &  1.065 &  0.103 &  0.974 &  0.044 &  \num{7.85d-04} \\
2  &  0.605 &  0.571 &  1.064 &  0.093 &  0.942 &  0.031 &  \num{3.46d-04} \\
3  &  0.596 &  0.553 &  1.073 &  0.082 &  0.927 &  0.025 &  \num{3.64d-04} \\
4  &  0.572 &  0.536 &  1.061 &  0.077 &  0.903 &  0.021 &  \num{3.06d-04} \\
5  &  0.571 &  0.538 &  1.039 &  0.071 &  0.927 &  0.019 &  \num{2.13d-04} \\
6  &  0.574 &  0.538 &  1.047 &  0.064 &  0.898 &  0.018 &  \num{2.13d-04} \\
7  &  0.571 &  0.554 &  1.051 &  0.059 &  0.878 &  0.017 &  \num{1.76d-04} \\
8  &  0.565 &  0.545 &  1.029 &  0.056 &  0.884 &  0.016 &  \num{1.47d-04} \\
9  &  0.551 &  0.544 &  0.990 &  0.055 &  0.907 &  0.014 &  \num{1.03d-04} \\
10 &  0.547 &  0.553 &  1.000 &  0.051 &  0.891 &  0.014 &  \num{1.05d-04} \\
      \bottomrule
  \end{tabular}
\end{table}

\section{Profile Correction}

The first remedy against the space-charge profile distortion is to increase the applied magnetic guiding field, such that the gyroradii are bounded by user defined limits as shown by \req{eq:fit-model}, \req{eq:fit-model-with-threshold} and \req{eq:fit-model-with-threshold-dependency} with corresponding parameters given by, respectively, \rtable{table:fit-results}, \rtable{table:fit-results-with-threshold} and \rtable{table:fit-results-with-threshold-dependency}. These relations are fit for the broad parameter range in \rtable{table:fit-region}. Most magnetic IPM designs fall into this category and further analysis is not required. However for some extreme scenarios very large magnetic fields are required~\cite{Sapinski-HB-2014}, which are both expensive to obtain and occupy significant space in already cramped synchrotrons. As technology advances more applications for high energy and high brightness beams are to be expected, putting current magnetic IPM designs to test. For these scenarios, correction mechanisms to obtain a measure of actual profile from the distorted profile, have been studied.

Foremost of them was the study of a hardware electron "sieve", which aims to filter electrons based on their gyroradius before they reach the acquisition system, and a numerical reconstruction on the resulting "sieved" profiles can be performed~\cite{Vilsmeier-Thesis}. Though the study with simulations gave promising results, the sieve was found to be rather complex and thick structure difficult to integrate in machine vacuum. Another approach was parametric curve fitting of the distorted profile with analytic functions and correlation of the beam width with those parameters~\cite{Sapinski-HB-2014}. The fit result however suffered from too few available data points and hence a general relationship between the fit parameters and the beam profile could not be established. 

Lately, approaches to record the inverse mapping between distorted profile to the original profile or the second moment of the original profile as a function of space charge parameters has been introduced either in form of look-up tables (LUT)~\cite{Jan-Egberts-Thesis} and supervised learning~\cite{IPAC-2018-ML}. All the aforementioned correction methods rely on well understood and benchmarked IPM simulations. In this section, the problem of profile correction is generally introduced.
Following that, several sub-approaches in the supervised learning scheme are mentioned and an extension to a full profile reconstruction from distorted profiles with arbitrary initial profile shapes is shown.

\subsection{Problem description}

In case of negligible space-charge interaction the distortion can be described via convolution of beam profile $P_{\textrm{beam}}$ with a \textit{point-spread function (PSF)} to obtain the measured profile $P_{\textrm{measured}}$~\cite{Sapinski-HB-2014}. This PSF depends on the initial velocities of electrons and can assessed by means of differential ionization cross sections. The PSF itself describes the electron transport in the IPM. It corresponds to the probability that an electron is detected on a certain position displaced from its ionization position and can be obtained from considering the time that an electron spends above the various bins of the detector~\cite{Vilsmeier-Thesis}. This PSF is independent of the position along the initial profile and hence the beam profile can be obtained by deconvolution.

If the PSF depends on the position along the profile - as in case of space-charge influence on electron movement - a convolution  cannot be used to describe the profile deformation anymore.
As a more general transformation a matrix multiplication can be used instead:
\begin{equation}\label{eq:matrix-formula}
 P_{\textrm{measured}, i} = M_{ij}\cdot P_{\textrm{beam}, j}
\end{equation}
where $M_{ij}$ is the probability that an electron which was generated at position $j$ is collected at position $i$.
Comparing with convolution for the space-charge free case, the matrix $M$ contains the PSF as columns, shifted across the rows with the PSF center at the diagonal ($M_{ij} = \textrm{PSF}[i - j]$).

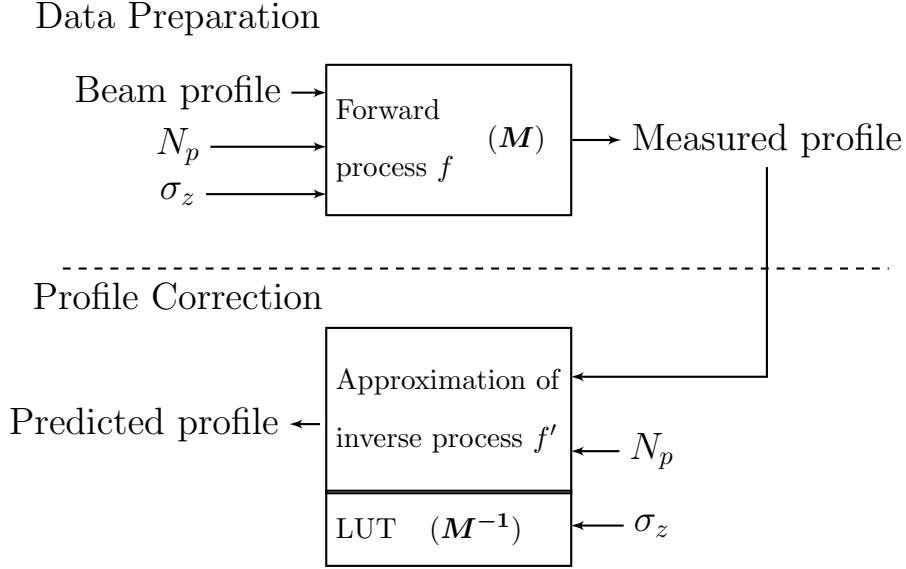
\begin{figure}[htbp]
\centering
\begin{tikzpicture}[scale=0.9]
\node[draw,thick,rectangle, minimum width = 3cm, minimum height = 2cm,text width = 3.0 cm](f) at (0,4){Forward process $f$};
\node[](dp) at (-4.0,5.8) {\large{Data Preparation}};
\node[](dp1) at (-4.0,1.7) {\large{Profile Correction}};
\node[](npi) at (-4.0,4.7) {\large{Beam profile}};
\node[](sxi) at (-4.0,3.9) {\large{$N_p$}};
\node[](sli) at (-4.0,3.2) {\large{$\sigma_z$}};
\node[](mpi) at (4.7,4.0) {\large{Measured profile}};
\node[](sxo) at (-4.5,-0.2) {\large{Predicted profile}};
\node[](npo) at (3.0,-0.6) {\large{$N_p$}};
\node[](slo) at (3.0,-1.7) {\large{$\sigma_z$}};
\node[draw,thick,rectangle,minimum width =3.1cm, minimum height = 1cm,text width = 3.0cm](f_in) at (0,-1.74){LUT };
\node[](M) at (1.0,4.0) {($\bm{M}$)};
\node[](min) at (0.4,-1.75) {($\bm{M^{-1}}$)};
\node[draw,thick,rectangle, minimum width =3.1cm, minimum height = 2.2cm,text width = 3.0 cm](f_in) at (0,0.0){Approximation of inverse process $f'$};
\draw[thick,dashed] (-5.7,2.1) --+ (0:12.2);
\draw[-latex',thick] (npi) --+ (0:2.2);
\draw[-latex',thick] (sxi) --+ (0:2.2);
\draw[-latex',thick] (sli) --+ (0:2.2);
\draw[-latex',thick] (slo) --+ (180:1.2);
\draw[latex'-,thick] (sxo) --+ (0:2.6);
\draw[-latex',thick] (npo) --+ (180:1.2);
\draw[-latex',thick] (f) --+ (mpi);
\draw[-latex',thick] (mpi) --++ (-90:3.5) --++(180:2.9);

\end{tikzpicture}
\caption{The physical process is modelled by a simulation tool and the inverse process is approximated by the chosen machine learning algorithm or stored parameter dependent inverse mapping matrix.}
\label{fig:model_inverse}
\end{figure}

The transformation matrix depends on the beam parameters $M_{ij}=M_{ij}(N_p, \sigma_x, \sigma_z)$ and for a given set of parameters
it can be obtained by simulations.
Performing simulations over a grid of space-charge distortion parameters one can establish a look-up table of matrices $M$ and their inverses $M^{-1}$. A fast iterative procedure to correct measured profiles using the simulated $M^{-1}$ matrices is discussed in \cite{Jan-Egberts-Thesis}. A newer approach of supervised machine learning is to deduce a set of rules for mapping measured profiles to their original counterparts. The idea is to infer a corresponding set of rules by providing distorted profiles alongside other relevant beam parameters and to map these to the original distribution of the beam. \rfig{fig:model_inverse} shows a schema of profile correction with the aid of simulation data.

\subsection{Data generation by IPM simulations}

A recent joint effort between laboratories led to the development of a generic simulation tool called "Virtual-IPM"~\cite{virtual-ipm}.
The Virtual-IPM tool has been used for simulating the movement of electrons inside the IPM, in presence of transient beam fields. \rtable{table:simulation-parameters} shows the parameter ranges that are used for the simulations in order to span the relevant parameter space.
The single bunches are modeled by three-dimensional Gaussian charge distributions. The electric field of bunches is computed via an analytic formula for a two-dimensional Gaussian charge distribution in the transverse plane~\cite{bassetti-erskine} while the longitudinal dependency is taken into account by rescaling the field with the beam's line density. The longitudinal field component is neglected. This approximation is justified because the beam is highly relativistic and the longitudinal dimension of bunches is significantly larger than their transverse dimensions and therefore the electric field is mainly acting in the transverse plane. The advantage of the analytic formula, as compared to a numerical Poisson solver, is that it is much faster in computing the electric field and it does not suffer from discretization effects.
Both the electric and the magnetic field of the beam are taken into account.
The external electric and magnetic guiding fields are modeled to be uniform within the field cage.
The initial velocities of electrons are generated according to a double differential cross section for a Hydrogen target~\cite{voitkiv}.
The passage of only a single bunch is simulated because the extraction times for electrons are only a few nanoseconds for the given electric guiding field while the bunch spacing is \SI{25}{\nano\second}.

\begin{table}[htbp]
  \centering
  \caption{Parameters for the simulation. The bunch population, length, width and height are varied, resulting in a total of \num{21021} data samples. The parameter ranges roughly correspond to the LHC flat top.}
  \label{table:simulation-parameters}
  \begin{tabular}{lcc}
      \toprule
         \textbf{Parameter}           & \textbf{Values} \\
      \midrule
         Beam particle type           & Protons \\
         Energy/u                     & \SI{6.5}{\tera\electronvolt} \\
         Bunch population             & \SIrange{1.1d11}{2.1d11}{ppb} \\
         Bunch length ($4\sigma_z$)   & \SIrange{0.9}{1.2}{\nano\second} \\
         Bunch width ($1\sigma_x$)    & \SIrange{270}{370}{\micro\metre} \\
         Bunch height ($1\sigma_y$)   & \SIrange{360}{600}{\micro\metre} \\
         Electrode distance           & \SI{85}{\milli\metre} \\
         Applied voltage              & \SI{4}{\kilo\volt} \\
         Magnetic field               & \SI{0.2}{\tesla} \\
         Number of sim. electrons     & \num{1000000} \\
         Time step size               & \SI{0.3125}{\pico\second} \\
      \bottomrule
  \end{tabular}
\end{table}

The output of the simulations is summarized by histograms with \SI{9.9}{\milli\meter} range and \SI{55}{\micro\meter} bin size, representing the electrons' positions at the moment of ionization, in the following referred to as \textit{initial}, and at the moment of detection, in the following referred to as \textit{final}. \SI{55}{\micro\meter} resolution corresponds to a hybrid-pixel detector type which has been successfully operated in the PS IPMs~\cite{hybrid-pixel-detector}.

\subsection{Reconstruction of beam profile standard deviation}

\req{eq:fit-model-with-threshold} has the interesting property that for a given measured standard deviation of $\sigma_m$ and a given magnetic field $B$ the formula encodes the corresponding beam profile standard deviation $\sigma_t$ which gives rise to the distortion via the threshold parameter $\tau = \sigma_m\sigma_t^{-1} - 1$. By rearranging \req{eq:fit-model-with-threshold} and inserting the definition of $\tau$ the beam profile standard deviation emerges as the root of the function

\begin{equation}
\label{eq:infer-beam-rms}
f(\sigma_t) = \frac{d\cdot N^a}{\sigma_z^b\sigma_t^c\left(\sigma_t^{-1}\sigma_m - 1\right)^g} + \frac{f}{\sigma_t^e\left(\sigma_t^{-1}\sigma_m - 1\right)^h} - B = 0
\end{equation}

on the interval $(0, \sigma_m)$, which can be computed by means of numerical methods.

\begin{figure}[htbp]
    \centering
    \includegraphics[width=8.6cm]{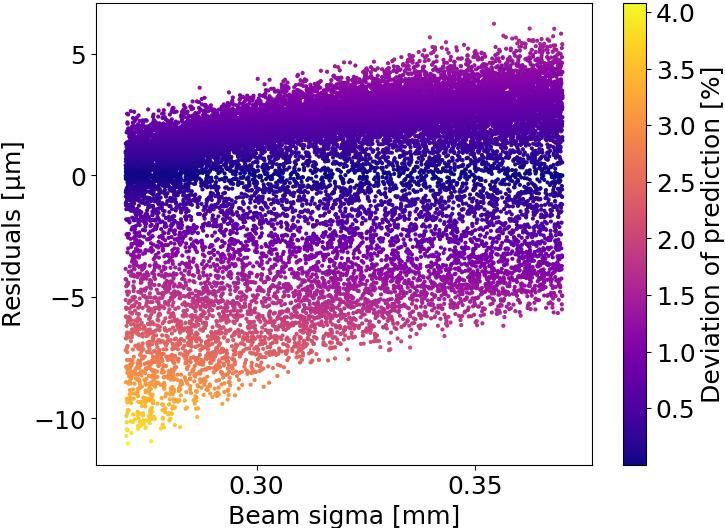}
    \caption{Residuals for the beam profile standard deviation reconstruction based on \req{eq:infer-beam-rms} and tested with simulations corresponding to \rtable{table:simulation-parameters}.}
    \label{fig:formula-based-rms-reconstruction-residuals-plot}
\end{figure}

We test this method on the simulation data prepared according to \rtable{table:simulation-parameters} by using bisection method~\cite{numerical-recipes} with \SI{0.1}{\micro\meter} tolerance. The resulting residuals have an overall mean and standard deviation of, respectively, \SI{0.357}{\micro\meter} and \SI{3.21}{\micro\meter}. The residuals plot \rfig{fig:formula-based-rms-reconstruction-residuals-plot} shows that the quality varies with beam profile standard deviation. Nevertheless most results are within 2\% accuracy. 

Previous attempts of reconstructing the initial beam profile standard deviation value from measured profiles using Machine Learning methods have shown a better performance while being fitted on a much smaller parameter range~\cite{IPAC-2018-ML}. \req{eq:fit-model-with-threshold} on the other hand has been fitted on a range which spans multiple orders of magnitude.

\subsection{Reconstruction of complete profiles with machine learning}

Relevant for the presented problem are methods to perform supervised regression for predicting continuous variables~\cite{elements-of-statistical-learning}. In general a supervised machine learning (ML) model represents an algorithm $f$, a mapping from input $x$ to output $y_p$, called \textit{decision function}, that is specified by a set of parameters $\theta \equiv \{\theta_i\}$. The inputs and outputs are exactly opposite to the data preparation stage as shown in Figure 8. While the structure of such an algorithm (e.g. the number of parameters $|\theta|$) depends on its hyper-parameters and is fixed, the goal is to tune $\theta_i$ such that the corresponding function $f(x\mid\theta)$ describes the output best.
The quality of this description is typically assessed by a so called \textit{loss function} $\mathbb{L}(x, y \mid \theta) \rightarrow \mathbb{R}^{+}$ which measures the deviation of the predictions $y_p$ from the target output values $y$.
Minimizing this loss function is used as a criterion for optimization of the model parameters $\theta$. In order to find a suitable ML algorithm for a given problem their performances are compared. For doing so the data set is split in three subsets corresponding to training, validation and testing. The training set is used to fit the particular ML algorithms in order to determine the optimal parameters $\theta$. The validation set is used for assessing the performance of an algorithm once it is fitted. Validating on a distinct data set prevents effects of over-fitting the algorithm to that particular training set and hence ensures generalization of the algorithm. The test similarly aims to prevent a tuning bias towards the validation set, resulting from the multiple iterations corresponding to hyperparameter tuning. The test set is used to assess the final performance that is to be expected once a ML algorithm with a specific configuration has been chosen.

\subsubsection{Full profile reconstruction}

Previous works have studied the usage of machine learning models for establishing a relationship between measured profile and original beam profile standard deviation~\cite{IBIC-2017-ML, IPAC-2018-ML}. A very recent approach investigated reconstruction of the complete profile shape from measured profiles~\cite{HB-2018-ML}. This approach used an additional global data transformation that is derived from the training data in order to converge during the fitting procedure. Here we present a novel approach that works only with per-profile normalization, providing a more consistent way of data preparation.

\paragraph{Model architecture}

\begin{figure}[htbp]
\centering
\begin{subfigure}[b]{0.45\textwidth}
 \centering
 \includegraphics[width=4.6cm, keepaspectratio]{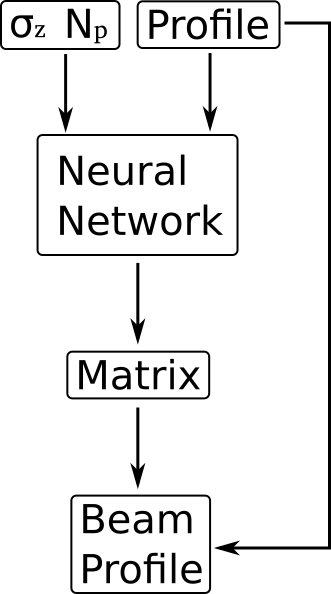}
 \caption{Profile reconstruction.}
 \label{fig:full-profile-sketch-ml-model-architecture}
\end{subfigure}
\quad
\begin{subfigure}[b]{0.45\textwidth}
 \centering
 \includegraphics[width=4.6cm, keepaspectratio]{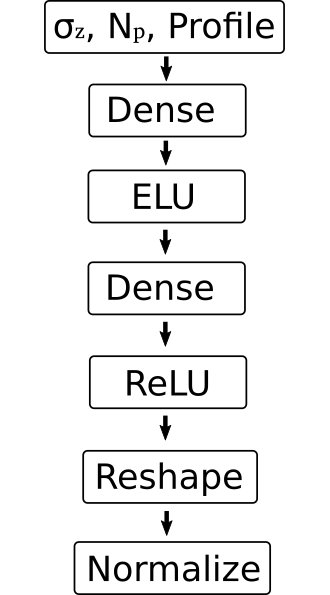}
 \caption{Matrix generation.}
 \label{fig:full-profile-sketch-feed-forward-example}
\end{subfigure}
\caption{Left: Sketch of the machine learning model which is used to perform the full profile reconstruction. The measured (distorted) profiles are used in two places, first to generate the transformation matrix and second for the subsequent dot product leading to the reconstructed beam profile. Right: Neural network architecture used for the matrix generation.}
\label{fig:full-profile-reconstruction-sketch}
\end{figure}

In order to establish a mapping between measured profiles together with bunch length and bunch intensity data to the original beam profiles we recall \req{eq:matrix-formula}. This equations states that the process of profile distortion can be described by a matrix multiplication $M$ of the beam profile $P_{\textrm{beam}}$. In general this matrix is dependent on the beam parameters and hence the beam profile itself: $M \rightarrow M(\sigma_z, N_p, P_{\textrm{beam}})$. We make the assumption that the matrix is invertible and that the inverse matrix $M^{-1}$ similarly depends on the measured profile: $M^{-1} \rightarrow M^{-1}(\sigma_z, N_p, P_{\textrm{measured}})$. Because the distortion arises from a symmetric process, the gyration of electrons, and the shift of final with respect to initial gyration center is negligible, as illustrated by \rfig{fig:example-trajectory} and discussed in Appendix \ref{section:gyrocenter-shift-compensation} and Ref. \cite{Vilsmeier-Thesis}, the columns of $M$ are linearly independent and hence the invertibility criterion is fulfilled. The second assumption of $M^{-1}$ being fully characterized by $P_{\mathrm{measured}}$ and $\sigma_z, N_p$ is less obvious but it is vital for the following algorithm. In that sense the algorithm will only be successful if that condition is fulfilled. In the following we use simulation data to verify the validity of this assumption. Given $M^{-1}(\sigma_z, N_p, P_{\textrm{measured}})$ the original profile can be obtained by means of a matrix multiplication $M^{-1}$ with the measured profile $P_{\textrm{measured}}$. Here we identify the task of the neural network as the generation of the inverse matrix in dependence of the measured profile as well as the additional beam parameters bunch length and bunch intensity. \rfig{fig:full-profile-reconstruction-sketch} sketches the procedure and shows the involvement of the machine learning model.
Since the final prediction is obtained via a matrix multiplication (linear transformation) the neural network can be regarded as a generator model which produces linear models.

\begin{figure}[htbp]
\centering
\includegraphics[width=8.6cm]{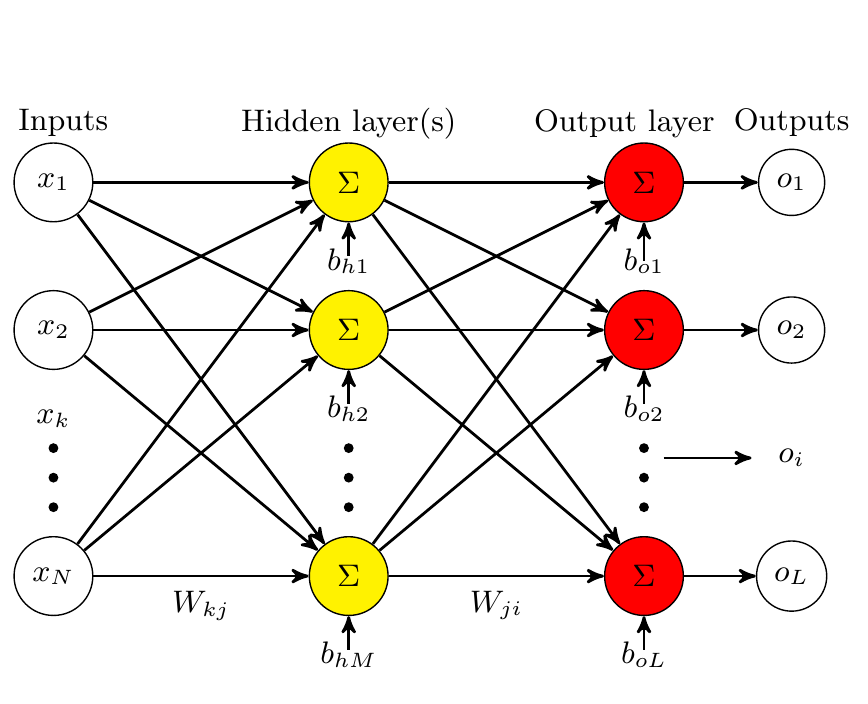}
\caption{Schematic of the architecture of a feed-forward multi-layer perceptron.}
\label{fig:MLP_ANN}
\end{figure}

\begin{figure}[htbp]
\centering
\includegraphics[width=8.6cm, keepaspectratio]{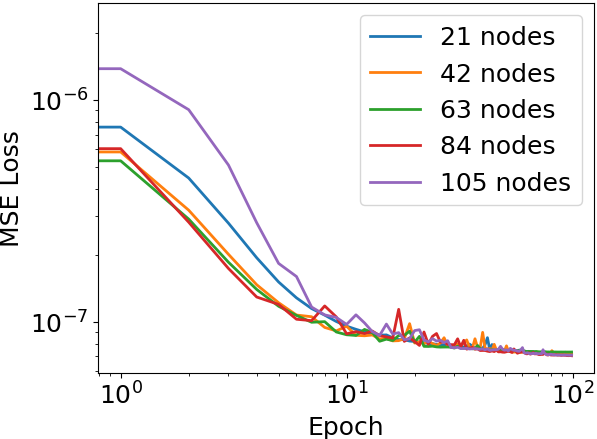}
\caption{Comparison of neural network performance for different widths of the first dense (hidden) layer. Other hyper-parameters were fixed to initializer = He-uniform, batch size = 236, learning rate = \num{8.882d-03}.}
\label{fig:nn-hidden-width-comparison}
\end{figure}

The architecture of the machine learning model and the neural network is depicted in \rfig{fig:full-profile-reconstruction-sketch}. The neural network uses two dense layers, that is layers which connect all nodes from the previous layer with their own nodes. The layers use ELU and ReLU activation function respectively. ELU stands for exponential linear unit and is defined by $\mathrm{ELU}(x) = \{x \; \mathrm{if} \; x \geq 0 \; \mathrm{else} \; e^x - 1\}$. ReLU stands for rectified linear unit and is defined by $\mathrm{ReLU}(x) = \max(0, x)$. Hence, up to this stage, the used architecture resembles a feed-forward network with a single hidden layer, as depicted in \rfig{fig:MLP_ANN}, where the output layer represents the flat transformation matrix. ReLU activation is used on that output layer because the physical interpretation of matrix elements being fractions of transferred electrons allows only entries greater than or equal to zero. The flat output is then reshaped to a square matrix and and each of its columns is normalized to unity due to the preservation of the electron signal.
We performed hyper-parameter search over the batch size, learning rate and the number of nodes in the first dense layer. The corresponding values are shown in \rtable{table:hyper-parameter-scan}. For identifying optimal configurations we monitored the final validation loss as well the ratio of initial to final validation loss in order to account for different starting losses. In a first step consensus about the batch size was found. With this optimal value fixed during a second and third iteration the optimal number of nodes and learning rate was identified respectively. The number of nodes however was found to have no significant influence on performance, if large enough, as shown in \rfig{fig:nn-hidden-width-comparison}. The final configuration was batch size 236 (evenly divides the training set of \num{13452} samples), learning rate \num{8.882d-03} and number of nodes in the first dense layer equal to 42. We used He-uniform initializer~\cite{He-initializer} and mean squared error (MSE) as the loss function.

\begin{table}[htbp]
  \centering
  \caption{Hyper-parameters that have been searched over and their values. Each hyper-parameter set is randomly sampled from the given distribution. Batch size is log-sampled with base 2 and learning rate and number of nodes are log-sampled with base 10. A total of \num{1000} samples were scanned over.}
  \label{table:hyper-parameter-scan}
  \begin{tabular}{lc}
      \toprule
          \textbf{Hyper-parameter} & \textbf{Range} \\
          \midrule
          Batch size & \numrange{1}{2048} \\
          Learning rate & \numrange{1d-2}{1d-6} \\
          Number of nodes & \numrange{10}{2288} \\
      \bottomrule
  \end{tabular}
\end{table}

\paragraph{Data preparation}

Data preparation is an important aspect of any machine learning analysis and helps the fitting procedure to converge. Also it needs to be ensured that the way of data preparation is compatible with real application requirements. The previous study on full profile reconstruction~\cite{HB-2018-ML} used per-profile normalization followed by a per-feature normalization which was computed over the full training set.
In the current study we focus on a more intuitive approach that only uses per-profile normalization.
For data preparation we use the following steps:
\begin{enumerate}
    \item Measured profiles are cropped to the range $[\SI{-3.90}{\milli\meter}, \SI{3.90}{\milli\meter}]$.
    \item Measured bunch length and number of charges per bunch are divided by, respectively, \SI{1.2}{\nano\second} and \num{2.1e11}.
    \item Measured profiles are normalized to unit integral.
    \item Measured profile bins are set to zero for values smaller than 5\% of the peak value.
\end{enumerate}

The previous approach on the other hand computed a global valid range by considering a 5\% peak-threshold region of the largest profile in the training set which was $[-\SI{0.93}{\milli\meter}, \SI{0.93}{\milli\meter}]$. All profile bins outside that region were dropped and not used for the analysis.
Hence, for comparison with the previous approach, we also apply this transformation in addition, by dropping data and predictions outside that valid range. However during training we use the full range available. 

\paragraph{Results}

We used mean squared error (MSE) as a loss function and monitored the mean absolute error (MAE) as an additional metric. For Gaussian profiles we also compare the deviation of standard deviation of the target beam profiles and predicted profiles. Fitting converged after 25 epochs, yielding mean and standard deviation of $\textrm{MSE} = \num{5.12d-07} \pm \num{2.49d-07}, \textrm{MAE} = \num{6.44d-04} \pm \num{1.70d-04}$ on the test data set. The deviation of resulting profile standard deviation was $\Delta\sigma_x / \sigma_x = \SI{2.01}{\percent} \pm \SI{1.00}{\percent}$ which is about one order of magnitude larger than what has been obtained from previous attempts with machine learning models directly targeting the profile standard deviation~\cite{IPAC-2018-ML}. \rtable{table:results-comparison-with-old-approach} shows the performance on the test set compared to the previously developed algorithm, the latter of which clearly performs better.

\begin{table}[htbp]
  \centering
  \caption{Performance of old and new approach on the test data set, containing \num{4205} samples.}
  \label{table:results-comparison-with-old-approach}
  \begin{tabular}{ccccc}
      \toprule
          \textbf{Loss} & \multicolumn{2}{c}{\textbf{MSE} [$10^{-7}$]} & \textbf{MAE} [$10^{-4}$] & \textbf{$\Delta\sigma/\sigma$ [\%]} \\
          \textbf{Method} & \textbf{meas.} & \textbf{rec.} & \textbf{rec.} & \textbf{rec.} \\
          \midrule
            Old & \multirow{ 2}{*}{$370 \pm 253$} & $0.32 \pm 0.11$ & $1.28 \pm 0.21$ & $0.15 \pm 0.11$ \\
            New &                     &$5.12 \pm 2.49$ & $6.44 \pm 1.70$ & $2.01 \pm 1.00$ \\
      \bottomrule
  \end{tabular}
\end{table}

The neural network has been trained on Gaussian profiles only. However in real world applications the actual beam shapes might deviate from this ideal scenario. Hence an important requirement is that the reconstruction algorithm still works even if applied to different beam shapes. In order to verify this generalization we tested the performance on various non-Gaussian shapes, sampled from Generalized Gaussian distributions and Q-Gaussian distributions. The charge distributions are given by the following equations.

Generalized Gaussian, used for instance in linac beam profile description \cite{Jan-Egberts-Thesis}: 
\begin{equation}\label{eq:generalized-gaussian}
    \frac{\beta}{2\alpha\Gamma(1/\beta)}\exp\left[-\left(\frac{\lvert x - \mu \rvert}{\alpha}\right)^{\beta}\right]
\end{equation}
with shape parameters $\alpha, \beta, \mu$; $\beta = 2, \alpha = \sqrt{2}\sigma_x$ corresponds to a normal distribution. $\Gamma$ denotes the gamma function~\cite{table-of-integrals}.

Q-Gaussian, used in studies of tails of hadron beams~\cite{q-gauss}: 
\begin{equation}\label{eq:q-gaussian}
    \frac{\sqrt{\beta}}{C_q}\left[1 - (1 - q)\beta x^2 \right]^{\frac{1}{1 - q}}
\end{equation}
with shape parameters $\beta, q$; $q = 1, \beta = \sigma_x^{-2} / 2$ corresponds to a normal distribution. $C_q$ is a normalization factor to provide unit integral.

\begin{figure}[htbp]
    \centering
    \begin{subfigure}[b]{0.45\textwidth}
        \centering
        \includegraphics[width=\textwidth]{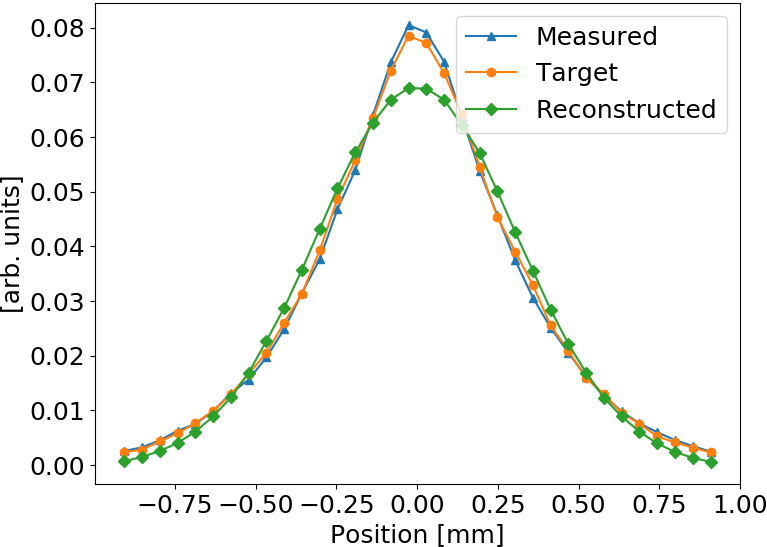}
        \caption[Generalized Gaussian $\beta = 1.5$]%
        {{\small Generalized Gaussian $\beta = 1.5$}}    
        \label{fig:example-profiles-gen-gauss-beta-1.5}
    \end{subfigure}
    \quad
    \begin{subfigure}[b]{0.45\textwidth}  
        \centering 
        \includegraphics[width=\textwidth]{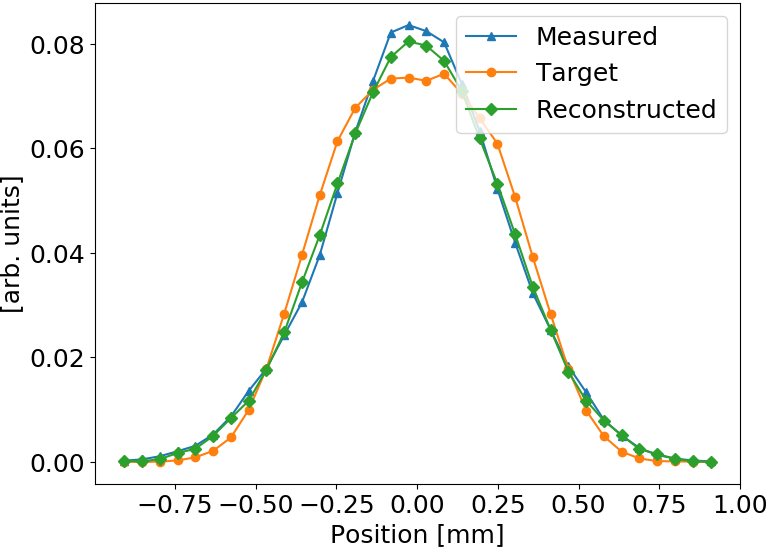}
        \caption[Generalized Gaussian $\beta = 3$]%
        {{\small Generalized Gaussian $\beta = 3$}}    
        \label{fig:example-profiles-gen-gauss-beta-3}
    \end{subfigure}
    \vskip\baselineskip
    \begin{subfigure}[b]{0.45\textwidth}   
        \centering 
        \includegraphics[width=\textwidth]{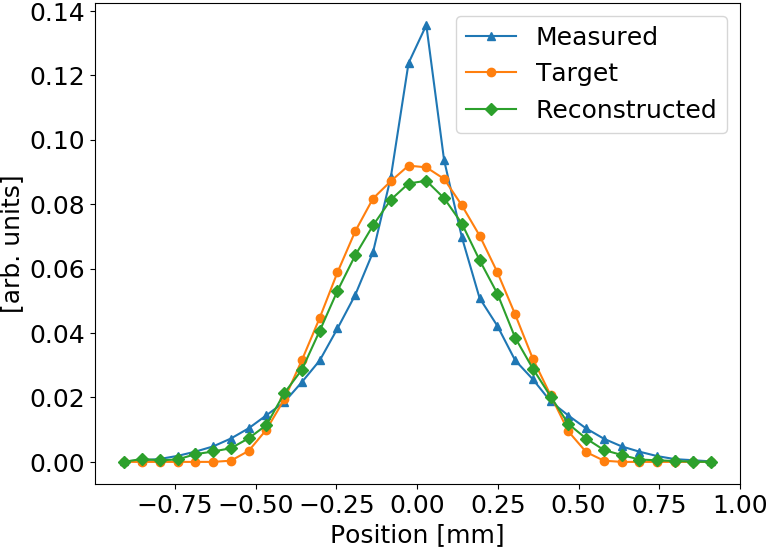}
        \caption[Q-Gaussian $q = 0.6$]%
        {{\small Q-Gaussian $q = 0.6$}}    
        \label{fig:example-profiles-q-gauss-q-0.6}
    \end{subfigure}
    \quad
    \begin{subfigure}[b]{0.45\textwidth}   
        \centering 
        \includegraphics[width=\textwidth]{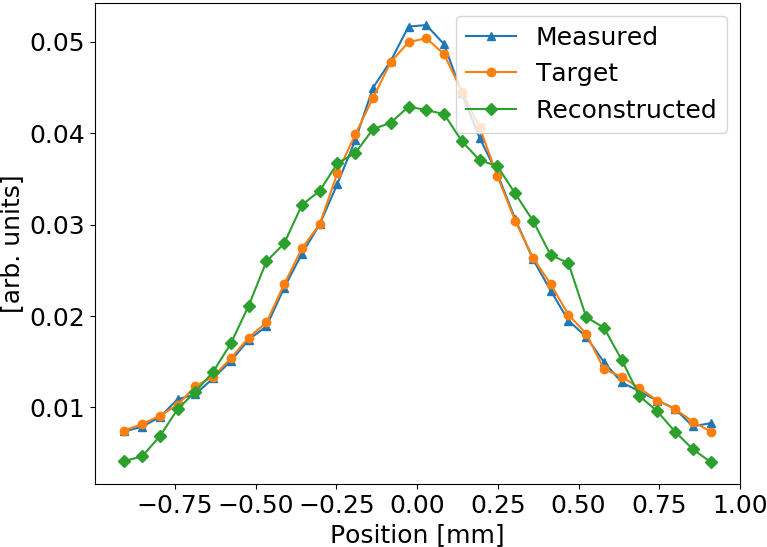}
        \caption[Q-Gaussian $q = 2$]%
        {{\small Q-Gaussian $q = 2$}}    
        \label{fig:example-profiles-q-gauss-q-2}
    \end{subfigure}
    \caption[Example profiles for non-Gaussian beam shapes]
    {Example profiles for non-Gaussian beam shapes together with reconstruction. ``Measured'' refers to the profile recorded by the detector (simulated in this case), ``Target'' refers to the initial beam distribution (the target of the reconstruction procedure) and ``Reconstructed'' refers to the result of the reconstruction procedure (which ideally should match the target). These examples are the ones with largest MSE error after reconstruction for each set. The Q-Gaussian profiles with $q = 2$ (bottom right corner) are significantly wider than the training profiles.} 
    \label{fig:example-profiles-non-Gaussian-beam-shapes}
\end{figure}

The parameters of the non-Gaussian shapes correspond to the ranges shown in \rtable{table:simulation-parameters} but are rescaled in order to match the Gaussian distribution in their limits. For the Generalized Gaussian $\alpha_{x, y, z} = \sigma_{x, y, z}$ and for the Q-Gaussian $\beta_{x, y, z} = \sigma_{x, y, z}^{-2} / 2$.
\rfig{fig:example-profiles-non-Gaussian-beam-shapes} shows an example profile for each profile shape together with the predicted reconstruction. Note that the Q-Gaussian profiles for $q = 2$ are significantly wider than the training profiles and as a result show only very little distortion.

\begin{figure}[htbp]
    \centering
    \begin{subfigure}[b]{0.45\textwidth}  
        \centering 
        \includegraphics[width=\textwidth]{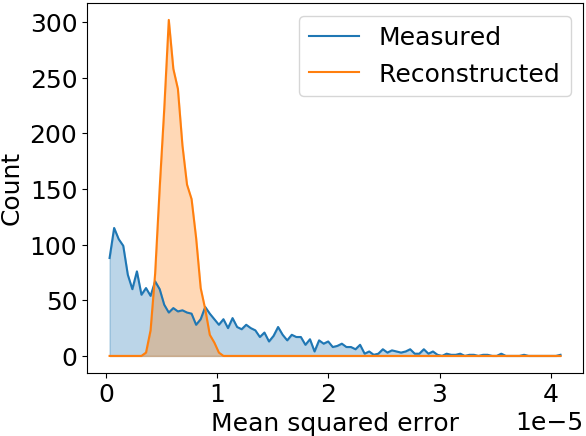}
        \caption[Generalized Gaussian $\beta = 1.5$]%
        {{\small Generalized Gaussian $\beta = 1.5$}}    
        \label{fig:mse-performance-gen-gauss-beta-1.5}
    \end{subfigure}
    \quad
    \begin{subfigure}[b]{0.45\textwidth}  
        \centering 
        \includegraphics[width=\textwidth]{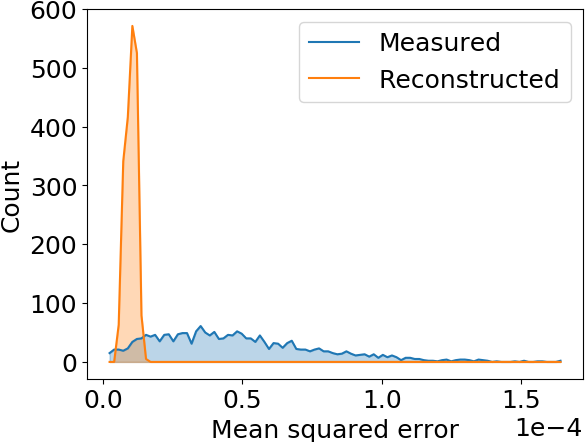}
        \caption[Generalized Gaussian $\beta = 3$]%
        {{\small Generalized Gaussian $\beta = 3$}}    
        \label{fig:mse-performance-gen-gauss-beta-3}
    \end{subfigure}
    \vskip\baselineskip
    \begin{subfigure}[b]{0.45\textwidth}  
        \centering 
        \includegraphics[width=\textwidth]{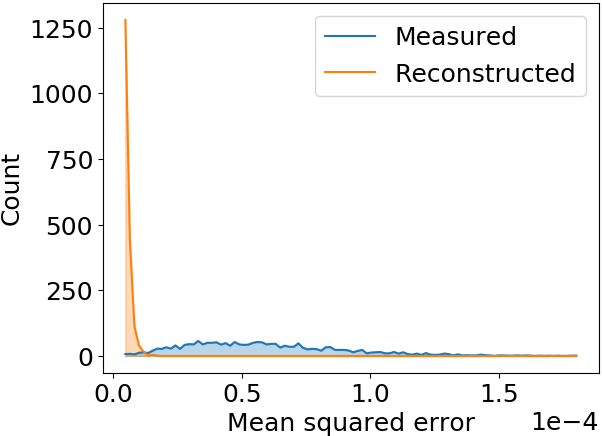}
        \caption[Q-Gaussian $q = 0.6$]%
        {{\small Q-Gaussian $q = 0.6$}}    
        \label{fig:mse-performance-q-gauss-q-0.6}
    \end{subfigure}
    \quad
    \begin{subfigure}[b]{0.45\textwidth}  
        \centering 
        \includegraphics[width=\textwidth]{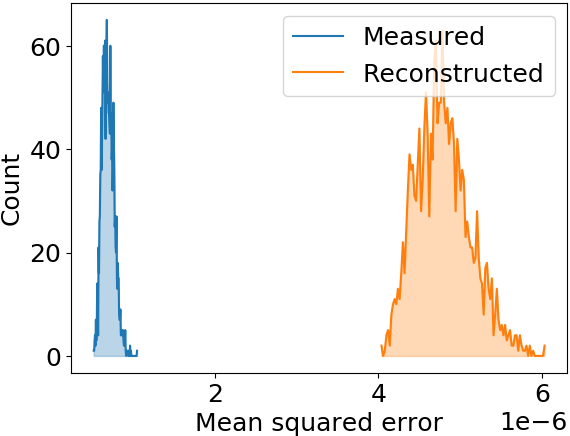}
        \caption[Q-Gaussian $q = 2$]%
        {{\small Q-Gaussian $q = 2$}}    
        \label{fig:mse-performance-q-gauss-q-2}
    \end{subfigure}
    \caption[Performance on non-Gaussian beam shapes]
    {Performance on non-Gaussian beam shapes. The Q-Gaussian profiles with $q = 2$ (bottom right corner) are significantly wider than the training profiles.} 
    \label{fig:mse-performance-non-Gaussian-beam-shapes}
\end{figure}

The overall performance, as shown in \rtable{table:performance-on-non-gaussian-beam-shapes} and \rfig{fig:mse-performance-non-Gaussian-beam-shapes}, decreases about one order of magnitude as compared to the Gaussian profiles. The new approach however shows improved performance when compared to the results of the previous approach. Because the wide $q = 2$ Q-Gaussian profiles significantly exceed the input range of the algorithm the corresponding profile tail information is dropped and hence the performance degrades. This is however not an issue as one can always validate the range of measured profiles before feeding them into the algorithm.
The obtained results on different profile shapes imply that the neural network identifies the underlying distortion mechanism rather than memorizing the specific profile shape which it has been trained on.

\begin{table}[htbp]
  \centering
  \caption{Test performance on various non-Gaussian beam shapes shown in \req{eq:generalized-gaussian} and \req{eq:q-gaussian}, measured by mean square error (MSE) in units of $10^{-6}$. The mean and standard deviation of MSE performance is given.}
  \label{table:performance-on-non-gaussian-beam-shapes}
  \begin{tabular}{ccccc}
      \toprule
          \textbf{Beam shape} & \textbf{Parameters} & \textbf{$\textrm{MSE}_{\textrm{meas}}$} & \textbf{$\textrm{MSE}_{\textrm{rec}}^{\textrm{old}}$} & \textbf{$\textrm{MSE}_{\textrm{rec}}^{\textrm{new}}$} \\
          \midrule
          Gen. Gaussian
          & $\beta = 1.5$ & $\num{8.09} \pm \num{6.92}$ & $\num{7.76} \pm \num{1.43}$ & $\num{6.57} \pm \num{1.20}$ \\
          & $\beta = 3$ & $\num{48.2} \pm \num{28.6}$ & $\num{14.5} \pm \num{1.27}$ & $\num{10.9} \pm \num{1.96}$ \\
          Q-Gaussian
          & $q = 0.6$ & $\num{59.8} \pm \num{30.4}$ & $\num{11.9} \pm \num{8.54}$ & $\num{6.30} \pm \num{1.47}$ \\
          & $q = 2$ & $\num{0.28} \pm \num{0.09}$ & $\num{105} \pm \num{28.9}$ & $\num{8.71} \pm \num{1.89}$ \\
      \bottomrule
  \end{tabular}
\end{table}

The results were obtained using various scientific Python packages, including Keras~\cite{keras}, Matplotlib~\cite{matplotlib}, NumPy~\cite{numpy}, Pandas~\cite{pandas}, SciPy~\cite{scipy}, Tensorflow~\cite{tensorflow} and Virtual-IPM~\cite{virtual-ipm}.

\section{Summary}

Distortions of measured beam profiles by electron-based magnetic Ionization Profile Monitors is a relatively new issue affecting the successful operation of these devices at some present and future accelerators operated for high-brightness beams.
We discussed the underlying effects and their magnitude for this type of distortion in detail at the example of a possible LHC flat top configuration, supported by both analytic calculations as well as complete numerical simulations.
Analytic descriptions for the simplified case of a uniform bunch shape are derived and benchmarking of these results with complete simulations was performed by using a well established simulation tool.
Simple formulae for computing the minimum magnetic field strength required to avoid space-charge induced profile distortion from relevant beam parameters are presented. They are validated against the derived analytic results for uniform beam as well as for complete simulations of various realistic beam shapes, including Uniform, Gaussian and Parabolic shapes.
Following that we addressed the problem of reconstruction of the real beam profile from distorted measurement. First an approach based on the previously developed formulas for minimum magnetic field is presented which allows to infer the beam profile standard deviation from measured profile standard deviation on a broad range of beam parameters.
The second approach addressed the problem of complete profile shape reconstruction and was realized by using a dedicated machine learning model involving neural network.
The performance of this method was tested on various realistic profile shapes, including Gaussian, Q-Gaussian and Generalized Gaussian shapes. By fitting the model only with Gaussian profiles we showed that the approach generalizes as well to beam shapes significantly different than normal distributions.


\newpage
\begin{appendices}

\section{Estimation of maximum polarization drift distance}
\label{section:estimation-of-polarization-drift-distance}

The polarization drift velocity is described by the formula~\cite{electromagnetic-drifts}

\begin{equation}
\label{eq:polarization-drift-velocity}
 v(t) = \frac{1}{\omega}\frac{d}{dt}\left(\frac{E_{\bot}}{B}\right)
\end{equation}

where $\omega$ is the gyrofrequency of electrons and $E_{\bot}$ is the electric field component perpendicular to the magnetic field $B$.

The maximum drift distance can be computed by

\begin{equation}
\label{eq:polarization-drift-distance}
 \delta = \int_0^{\infty}v(t)\,dt
\end{equation}

where $t = 0$ describes the time when the electron is longitudinally aligned with the bunch center, since for $t < 0$ the electron drifts in the opposite direction. For estimating the bunch electric field we use the formula for the transverse electric field of a two-dimensional Gaussian bunch scaled with the longitudinal charge line density:

\begin{equation}
\label{eq:symmetric-gaussian-electric-field}
 E_r(x, y, t) = \frac{Ne}{\epsilon_0}\frac{1}{\sqrt{2\pi}^3\sigma_z}\left[1 - \exp\left(-\frac{x^2 + y^2}{2\sigma_r^2}\right)\right]\exp\left(-\frac{\beta^2 c^2 t^2}{2\sigma_z^2}\right)
\end{equation}

where $N$ is the number of charges $\sigma_r$ is the standard deviation of the transverse charge distribution and $\sigma_z$ is the standard deviation of the longitudinal charge line density (in dimension of length); $\beta$ is the relativistic beta factor of the particle beam.

We further neglect the time dependence of $x$ and $y$ because we consider $\beta \approx 1$ together with non-relativistic electrons and hence the change in beam electric field will be driven by the beam movement which is expressed through the explicit time dependence in equation \req{eq:symmetric-gaussian-electric-field}. The beam magnetic field is neglected as well since it is much smaller than the magnetic guiding field. Hence the electric field component $E_{\bot}$ is determined to be perpendicular to the magnetic guiding field and is thus given by \req{eq:symmetric-gaussian-electric-field}.

It follows that the time derivative in \req{eq:polarization-drift-velocity} is given by the partial time derivative of \req{eq:symmetric-gaussian-electric-field}. Integrating the resulting expression according to \req{eq:polarization-drift-distance} one obtains (dropping the sign):

\begin{equation}
 \delta(x, y) = \frac{1}{\omega B}\frac{Ne}{\epsilon_0}\frac{1}{\sqrt{2\pi}^3\sigma_z}\left[1 - \exp\left(-\frac{x^2 + y^2}{2\sigma_r^2}\right)\right]
\end{equation}

The maximum of the resulting function $\delta(x)$ is given by:

\begin{equation}
\begin{aligned}
 \delta_{max} &= \frac{1}{\omega B}\frac{N}{\sigma_r\sigma_z}\frac{e}{\epsilon_0\sqrt{2\pi}^3}\frac{1}{\sqrt{-\bar{w}}}\left[1 - \exp\left(\frac{\bar{w}}{2}\right)\right] \\
              & \approx \frac{1}{\omega B}\frac{N}{\sigma_r\sigma_z}\num{5.18d-10}
\end{aligned}
\end{equation}

where $\bar{w} = 2w_{-1}\left(-0.5\exp(-0.5)\right) + 1$ and $w_{-1}$ is the product log function~\cite{table-of-integrals}.

Inserting the the beam parameters from \rtable{table:example-case-parameters} and using $\sigma_r = \sigma_x$ we obtain a maximum drift distance of $\delta_{max} = \SI{852}{\micro\meter}$.

\section{Compensation of gyration center shift}
\label{section:gyrocenter-shift-compensation}

\req{eq:shifted-gyrocenter} indicates the shift of the gyration center with respect to the undisturbed case (for which $E = 0$). Using \req{eq:definitions} this becomes:
\begin{equation}
    x_c = \frac{\omega^2}{\Omega^2}\left(x_0 - \frac{v_{z0}}{\omega}\right)
\end{equation}
with $\omega = \bar{q}B$ the gyrofrequency for the undisturbed case $E = 0$.

For the derivation of \req{eq:uniform-bunch-motion-x} -- \req{eq:uniform-bunch-motion-vz} the bunch electric field was assumed constant, i.e. $E = \textrm{const}$. In order to incorporate the time-dependence due to bunch movement we can consider a series of partial solutions $(x_n(t), z_n(t))$ each of which is valid on an interval $[t_n, t_{n+1}]$ for which the time dependence remains approximately constant. These partial solutions are of the form:

\begin{equation}\label{eq:partial-solutions}
\begin{aligned}
x_n(t) &= \frac{a_n}{\Omega_n}\sin(\Omega_n t) - \frac{b_n}{\Omega_n}\cos(\Omega_n t) + c_n \\
\dot{x}_n(t) &= a_n\cos(\Omega_n t) + b_n\sin(\Omega_n t) \\
z_n(t) &= -a_n\frac{\omega}{\Omega_n^2}\cos(\Omega_n t) - b_n\frac{\omega}{\Omega_n^2}\sin(\Omega_n t) + \frac{E_n}{B}c_n\,t + d_n \\
\dot{z}_n(t) &= a_n\frac{\omega}{\Omega_n}\sin(\Omega_n t) - b_n\frac{\omega}{\Omega_n}\cos(\Omega_n t) + \frac{E_n}{B}c_n
\end{aligned}
\end{equation}

where $E_n \equiv E(t_n)$ and

\begin{equation}
\begin{aligned}
a_0 &= v_{x0} \\
b_0 &= -\frac{\omega}{\Omega_0}\left(v_{z0} - \frac{E_0}{B}x_0\right) \\
c_0 &= \frac{\omega^2}{\Omega_0^2}\left(x_0 - \frac{v_{z0}}{\omega}\right) \\
d_0 &= z_0 + \frac{\omega}{\Omega_0^2}v_{x0}
\end{aligned}
\end{equation}
Note that $c_0$ represents the (shifted) gyration center.

In order to link the solutions and their coefficients together we require continuity at the boundary of time intervals:

\begin{equation}\label{eq:continuity-condition}
\left(x_n(t_n), \dot{x}_n(t_n), z_n(t_n), \dot{z}_n(t_n)\right) \stackrel{!}{=} \left(x_{n-1}(t_n), \dot{x}_{n-1}(t_n), z_{n-1}(t_n), \dot{z}_{n-1}(t_n)\right)
\end{equation}

Using \req{eq:partial-solutions} we can write \req{eq:continuity-condition} in matrix representation with respect to the coefficients:

\begin{equation}
\begin{pmatrix}
\frac{\sin(\Omega_n t_n)}{\Omega_n} & -\frac{\cos(\Omega_n t_n)}{\Omega_n} & 1 & 0 \\
\cos(\Omega_n t_n) & \sin(\Omega_n t_n) & 0 & 0 \\
-\frac{\omega}{\Omega_n^2}\cos(\Omega_n t_n) & -\frac{\omega}{\Omega_n^2}\sin(\Omega_n t_n) & \frac{E_n}{B}t_n & 1 \\
\frac{\omega}{\Omega_n}\sin(\Omega_n t_n) & -\frac{\omega}{\Omega_n}\cos(\Omega_n t_n) & \frac{E_n}{B} & 0
\end{pmatrix}
\begin{pmatrix} a_n \\ b_n \\ c_n \\ d_n \end{pmatrix}
= \begin{pmatrix} x_{n-1}(t_n) \\ \dot{x}_{n-1}(t_n) \\ z_{n-1}(t_n) \\ \dot{z}_{n-1}(t_n) \end{pmatrix}
\end{equation}

The matrix on the left hand side has determinant $-\Omega\omega^{-1} \neq 0$ and hence a unique solution for the coefficients $(a_n, b_n, c_n, d_n)$ exists. Inverting the matrix we obtain:

\begin{equation}
\begin{pmatrix}
\frac{\Omega_n^2 - \omega^2}{\Omega_n}\sin(\Omega_n t_n) & \cos(\Omega_n t_n) & 0 & \frac{\omega}{\Omega_n}\sin(\Omega_n t_n) \\
\frac{\omega^2 - \Omega_n^2}{\Omega_n}\cos(\Omega_n t_n) & \sin(\Omega_n t_n) & 0 & -\frac{\omega}{\Omega_n}\cos(\Omega_n t_n) \\
\frac{\omega^2}{\Omega_n^2} & 0 & 0 & -\frac{\omega}{\Omega_n} \\
\frac{\omega\Omega_n^2 - \omega^3}{\Omega_n^2}t_n & \frac{\omega}{\Omega_n^2} & 1 & \frac{\omega^2 - \Omega_n^2}{\Omega_n^2}t_n
\end{pmatrix}
\end{equation}

The parameter $c_n$ encodes the gyration center shift and is given by the third line of the inverse matrix:
\begin{equation}\label{eq:cn}
    c_n = \frac{\omega^2}{\Omega_n^2}x_{n-1}(t_n) - \frac{\omega}{\Omega_n^2}\dot{z}_{n-1}(t_n)
\end{equation}

Inserting \req{eq:partial-solutions} into \req{eq:cn} we obtain:
\begin{equation}
    c_n = \frac{\Omega_{n-1}^2}{\Omega_n^2}c_{n-1} = c_0\prod_{i=0}^{n-1}\frac{\Omega_i^2}{\Omega_{i+1}^2} = \frac{\Omega_0^2}{\Omega_n^2}c_0 = \frac{\omega^2}{\Omega_n^2}\left(x_0 - \frac{v_{z0}}{\omega}\right)
\end{equation}

Since for $n \rightarrow \infty$ we have $\Omega_n \rightarrow \omega$ it follows that $c_n \rightarrow x_0 - v_{z0}\omega^{-1}$ which is the undisturbed gyration center. Hence the initial shift, encoded in $c_0$, is compensated for $t \rightarrow \infty$. Actually this limit is already reached for $E \rightarrow 0$ which is the case once the bunch has receded from the electron's position.

\section{Derivation of time-of-flight for uniform charge distribution}
\label{section:tof-for-uniform-shape}

We define $a \equiv \frac{qE_g}{m},\;b \equiv \frac{q\lvert E\rvert}{m}$ and thus obtain:
\begin{equation}
    \ddot{y}(t) = a - by\exp\left(-\frac{\lvert t\rvert}{\sigma_z}\right)
\end{equation}

We solve the differential equation for $t \geq 0$ (denoted as $y^{>}$) since for $t \leq 0$ (denoted as $y^{<}$) we can then reuse this solution together with the transformation $t \mapsto -t$.

For $t \geq 0$ the solution is:
\begin{equation}
\label{eq:uniform-bunch-motion-y}
    \begin{aligned}
        y(t) &= \eta\sigma_z\J{0}{u(t)} G_{2,4}^{3,0}\left\{\frac{u(t)}{2}\right\} \\
             &- \eta\sigma_z\Y{0}{u(t)} G_{2,0}^{1,3}\left\{\frac{u(t)^2}{4}\right\} \\
             &+ k_1\J{0}{u(t)} + k_2\Y{0}{u(t)}
    \end{aligned}
\end{equation}
where $\eta \equiv a\sigma_z\pi$, $u(t) \equiv 2\sigma_z\sqrt{b}\exp^{\frac{-t}{2\sigma_z}}$, $J_n$, $Y_n$ denote the Bessel functions of first and second kind respectively and $G_{2,4}^{3,0}\left\{x\right\} \equiv G_{2,4}^{3,0}\left\{x, \frac{1}{2} \left| \substack{-\frac{1}{2}, 1 \\ 0, 0, 0, -\frac{1}{2}}\right.\right\}$ and $G_{2,0}^{1,3}\left\{x\right\} \equiv G_{2,0}^{1,3}\left\{x \left| \substack{1 \\ 0, 0, 0}\right.\right\}$ denote Meijer G-functions~\cite{table-of-integrals}.

By using the relations
\begin{equation}
    \begin{aligned}
        \frac{d}{dx}G_{2,4}^{3,0}\left\{x\right\} &= -\frac{\J{0}{2\sqrt{x}}}{x} \\
        \frac{d}{dx}G_{2,0}^{1,3}\left\{x\right\} &= -\frac{2\Y{0}{2x}}{x}
    \end{aligned}
\end{equation}
we confirm that the terms which contain the derivatives of the Meijer G-function cancel each other.
We then obtain for $\dot{y}(t)$:
\begin{equation}
\label{eq:uniform-bunch-motion-vy}
    \begin{aligned}
        \dot{y}(t) = -\dot{u}(t)[ & \eta\sigma_z\J{1}{u(t)}G_{2,4}^{3,0}\left\{\frac{u(t)}{2}\right\} \\
                                  & - \eta\sigma_z\Y{1}{u(t)}G_{2,0}^{1,3}\left\{\frac{u(t)^2}{4}\right\} \\
                                  & + k_1\J{1}{u(t)} + k_2\Y{1}{u(t)}]
    \end{aligned}
\end{equation}

Using the initial conditions $y(0) \stackrel{!}{=} y_0$ and $\dot{y}(0) \stackrel{!}{=} v_0$ we can derive the values for $k_1, k_2$ (setting $u_0 \equiv u(t_0)$)
\begin{equation}
    \begin{pmatrix}
        \J{0}{u_0} & \Y{0}{u_0} \\
        \J{1}{u_0} & \Y{1}{u_0}
    \end{pmatrix}
    \begin{pmatrix}
        k_1 \\
        k_2
    \end{pmatrix}
    =
    \begin{pmatrix}
        y_0 - \eta\J{0}{u_0}G_{2,4}^{3,0}\left\{\frac{u_0}{2}\right\} + \eta\Y{0}{u_0}G_{2,0}^{1,3}\left\{\frac{u_0^2}{4}\right\} \\
        \frac{-v_{y0}}{\dot{u}(0)} - \eta\J{1}{u_0}G_{2,4}^{3,0}\left\{\frac{u_0}{2}\right\} + \eta\Y{1}{u_0}G_{2,0}^{1,3}\left\{\frac{u_0^2}{4}\right\} \\
    \end{pmatrix}
\end{equation}

For the $t \leq 0$ case we apply $t \mapsto -t$ and reuse the above solution. Note that the derivative $\dot{u}(t)$ changes its sign due to the transformation. For $t_0 < 0$ we obtain $k^{>}_1, k^{>}_2$ from the continuity condition at $t = 0$: $y^{<}(0) \stackrel{!}{=} y^{>}(0)$ and $\dot{y}^{<}(0) \stackrel{!}{=} \dot{y}^{>}(0)$).

\section{Methods for minimum magnetic field computation}\label{sec:appendix:methods}

For the computation of minimum magnetic field strength from analytic formulae we used the following methods. For each configuration (beam parameters + IPM parameters including magnetic field) \num{1000} particles were sampled from the initial beam distribution (sampling x- and y-position as well as the ionization time at $z = 0$ since the guiding fields are assumed to be uniform and the resulting profile is integrated along z). The initial momenta are sampled from double differential cross sections~\cite{voitkiv}. For each particle its time-of-flight (TOF) until leaving the bunch region, described by $x(t)^2 + y(t)^2 \leq R^2 \;\,\textrm{and}\;\, z(t) \leq (t - 4\sigma_z)\beta c$, was computed as the minimum between the TOF in transverse direction, considering \req{eq:uniform-bunch-motion-x} and \req{eq:uniform-bunch-motion-y}, as well as in z-direction, using the bunch velocity and the electron velocity in z-direction via \req{eq:uniform-bunch-motion-z} and \req{eq:uniform-bunch-motion-vz}. We then use \req{eq:uniform-bunch-motion-vx} and \req{eq:uniform-bunch-motion-vz} in order to compute the final transverse velocity $\lvert\bm{v_{xz}}(t_{\textrm{of}})\rvert = \sqrt{v_x(t_{\textrm{of}})^2 + v_z(t_{\textrm{of}})^2}$ from which we calculate the gyroradius given the magnetic guiding field. We then consider gyration around the central point indicated by \req{eq:uniform-bunch-motion-x} and a random position of detection within the gyromotion range with probabilities proportional to the corresponding point-spread function (PSF)~\cite{Vilsmeier-Thesis}. In order to increase the statistics, for each particle \num{100} final positions were sampled according to the corresponding PSF, all corresponding to the same initial position. This results in a total of \num{100000} data points for computing the standard deviation of the beam profile and the measured profile respectively. Each configuration was scored according to the relative deviation of measured standard deviation with respect to standard deviation of the beam profile $\lambda = (\sigma_{m} - \sigma_b)\sigma_b^{-1}$, where $\lambda$ denotes the score and $\sigma_m, \sigma_b$ denote the measured and beam profile standard deviation respectively. A particular magnetic field strength is considered sufficient if $\lambda \leq 0.01$. The magnetic field strength $B$ is computed by finding the root of the function $\lambda(B) - 0.01$ using the bisection method~\cite{numerical-recipes} with a tolerance of \SI{1}{\milli\tesla}.

\section{Scaling factors for different bunch shapes}
\label{section:bunch-shape-scaling-factors}

Different bunch shapes are compared to a Gaussian distribution and their free parameters are adjusted such that the resulting distribution has a minimal least square difference to the corresponding Gaussian distribution:

\begin{equation}\label{eq:p-norm-difference-bunch-shapes}
    \argmin_{\eta} \int_{-\infty}^{+\infty}\left(\rho(x|\eta) - \mathcal{N}(0,\sigma)\right)^p\,dx
\end{equation}

where $\rho(x|\eta)$ represents a particular bunch shape parametrized by $\eta$. Least squares difference ($p = 2$) is used for the Uniform bunch shape and $p = 4$  is used for the Parabolic Ellipsoid shape for numerical stability. The L-BFGS-B solver from the scipy.optimize package~\cite{scipy} is used for minimizing \req{eq:p-norm-difference-bunch-shapes}.

\subsection{Gaussian}

The reference beam profile is a Gaussian distribution given by:
\begin{equation}
    \rho_G(x) \propto \exp\left(-\frac{x^2}{\sigma_t^2}\right)
\end{equation}

\subsection{Uniform}

We consider a charge distribution that is uniform in the transverse shape and has a Gaussian dependency in the longitudinal direction:
\begin{equation}
    \rho_U(x, y, z) \propto \begin{cases}
        \exp\left(-\frac{z^2}{2\sigma_z^2}\right) &, \textrm{for}\;\; x^2 + y^2 \leq R^2 \\
        0 &, \textrm{otherwise}
    \end{cases}
\end{equation}

Integration over $z$ and $y$ yields:
\begin{equation}
    \rho_U(x|R) \propto \sqrt{R^2 - x^2}
\end{equation}

\subsection{Parabolic Ellipsoid}

We consider a rotational symmetric ellipsoid with parabolic charge distribution:
\begin{equation}
    \rho_P(x, y, z) \propto \begin{cases}
        1 - \frac{x^2 + y^2}{b} - \frac{z^2}{a^2} &, \textrm{for}\;\; \frac{x^2 + y^2}{b^2} + \frac{z^2}{a^2} \leq 1\\
        0 &, \textrm{otherwise}
    \end{cases}
\end{equation}

Integration over $z$ and $y$ yields:
\begin{equation}
    \rho_P(x|\{a, b\}) \propto \frac{a}{b^3}\left(b^2 - x^2\right)^2
\end{equation}

\end{appendices}

\end{document}